\documentclass[sigconf]{acmart}

\AtBeginDocument{%
  }

\setcopyright{acmlicensed}
\copyrightyear{2018}
\acmYear{2018}
\acmDOI{XXXXXXX.XXXXXXX}
\acmConference[Conference acronym 'XX]{Make sure to enter the correct
  conference title from your rights confirmation email}{June 03--05,
  2018}{Woodstock, NY}
\acmISBN{978-1-4503-XXXX-X/2018/06}

\usepackage{multirow}
\usepackage{soul}
\usepackage{cleveref}
\usepackage{listings}
\usepackage{xcolor}
\usepackage{xspace}
\usepackage{tikz}
\usepackage{enumitem}
\usepackage{makecell}
\usepackage{booktabs}
\usepackage{tabularx}
\usepackage[inkscapelatex=false]{svg}
\usepackage{subcaption}
\usepackage{fancyvrb}
\usepackage{xcolor}

\DefineVerbatimEnvironment{CodeBox}{Verbatim}{
    frame=single,
    framesep=5pt,
    fontsize=\small,
    rulecolor=\color{black!30},
    breaklines=true,
    breakanywhere=true
}

\lstdefinestyle{codebox}{
    frame=single,
    framesep=5pt,
    basicstyle=\ttfamily\small,
    backgroundcolor=\color{gray!10},
    rulecolor=\color{black!30},
    moredelim=[s][\color{blue}]{\{state}{\}},
    moredelim=[s][\color{blue}]{\{TIME}{\}},
    moredelim=[s][\color{blue}]{\{VISUALIZATION}{\}},
    moredelim=[s][\color{blue}]{\{TEMPLATE_CODE}{\}},
    moredelim=[s][\color{blue}]{\{vis}{\}},
    breaklines=true,
    breakatwhitespace=true,
    breakindent=0pt,           
    showstringspaces=false,
    columns=flexible,
    xleftmargin=0pt,          
    resetmargins=true         
}

\usetikzlibrary{shadings}

\definecolor{cback}{HTML}{E9ECEF}
\definecolor{cframe}{HTML}{495057}
\definecolor{clightblue}{HTML}{DDEFFC}
\definecolor{clightgreen}{HTML}{DFFFE5}
\definecolor{clightyellow}{HTML}{FFF4DB}

\definecolor{cback1}{HTML}{F8F9FA} 
\definecolor{cframe1}{HTML}{ADB5BD} 

\definecolor{cback2}{HTML}{F1F3F5} 
\definecolor{cframe2}{HTML}{CED4DA} 

\definecolor{cback3}{HTML}{EBF5FB} 
\definecolor{cframe3}{HTML}{A9CCE3} 

\definecolor{cback4}{HTML}{FDF3E7} 
\definecolor{cframe4}{HTML}{F5CBA7} 

\newcommand*\circled[1]{\tikz[baseline=(char.base)]{
    \node[shape=rectangle,rounded corners=1.5pt,fill=cback,text=black,draw=cframe,inner sep=1pt] (char) {#1};}}

\newcommand*\circledVarD[1]{\tikz[baseline=(char.base)]{
    \node[shape=rectangle,rounded corners=1.5pt,fill=cback4,text=black,draw=cframe4,inner sep=1pt] (char) {#1};}}

\newcommand{\sys}{\textcolor{black}{\textsc{Augmented Algorithm}}\xspace}

\begin{document}

\title{Augmented Algorithm: Generating Interactive Algorithm Visualizations from Static Textbook Pseudocode}

\author{Yuki Ueno}
\orcid{1234-5678-9012}
\affiliation{%
  \institution{Arizona State University}
  \city{Tempe}
  \state{Arizona}
  \country{USA}
}
\email{yueno@asu.edu}

\author{Zhuojun Jiang}
\affiliation{%
  \institution{Arizona State University}
  \city{Tempe}
  \state{Arizona}
  \country{USA}
}
\email{zjian115@asu.edu}

\author{Chris Bryan}
\affiliation{%
  \institution{Arizona State University}
  \city{Tempe}
  \state{Arizona}
  \country{USA}
}
\email{cbryan16@asu.edu}



\begin{abstract}
Learning data structures and algorithms is a fundamental activity in computer science education, but this is challenging due to their abstract and dynamic nature. While algorithm visualization tools have shown pedagogical benefits, they remain underutilized in educational settings due to creation or integration difficulties. We introduce \sys, a novel LLM-powered algorithm visualization tool that transforms static textbook pseudocode into embedded, interactive algorithm visualizations. Our web-based system combines computer vision with LLMs to automatically generate synchronized, step-by-step algorithm animations from pseudocode directly within scanned textbook pages. We evaluated \sys through technical evaluation, a usability study with students (N=15), and expert interviews with instructors (N=6). Results show that our system significantly reduces algorithm visualization development and access effort while enabling more interactive and tailored learning experiences for algorithmic concepts. All supplemental materials required to reproduce the study are available on \url{https://osf.io/5xrmy/overview?view_only=6954b8873f5e404e890aee91ab1863ac}.
\end{abstract}

\begin{teaserfigure}
  \centering
  \includegraphics[width=\textwidth]{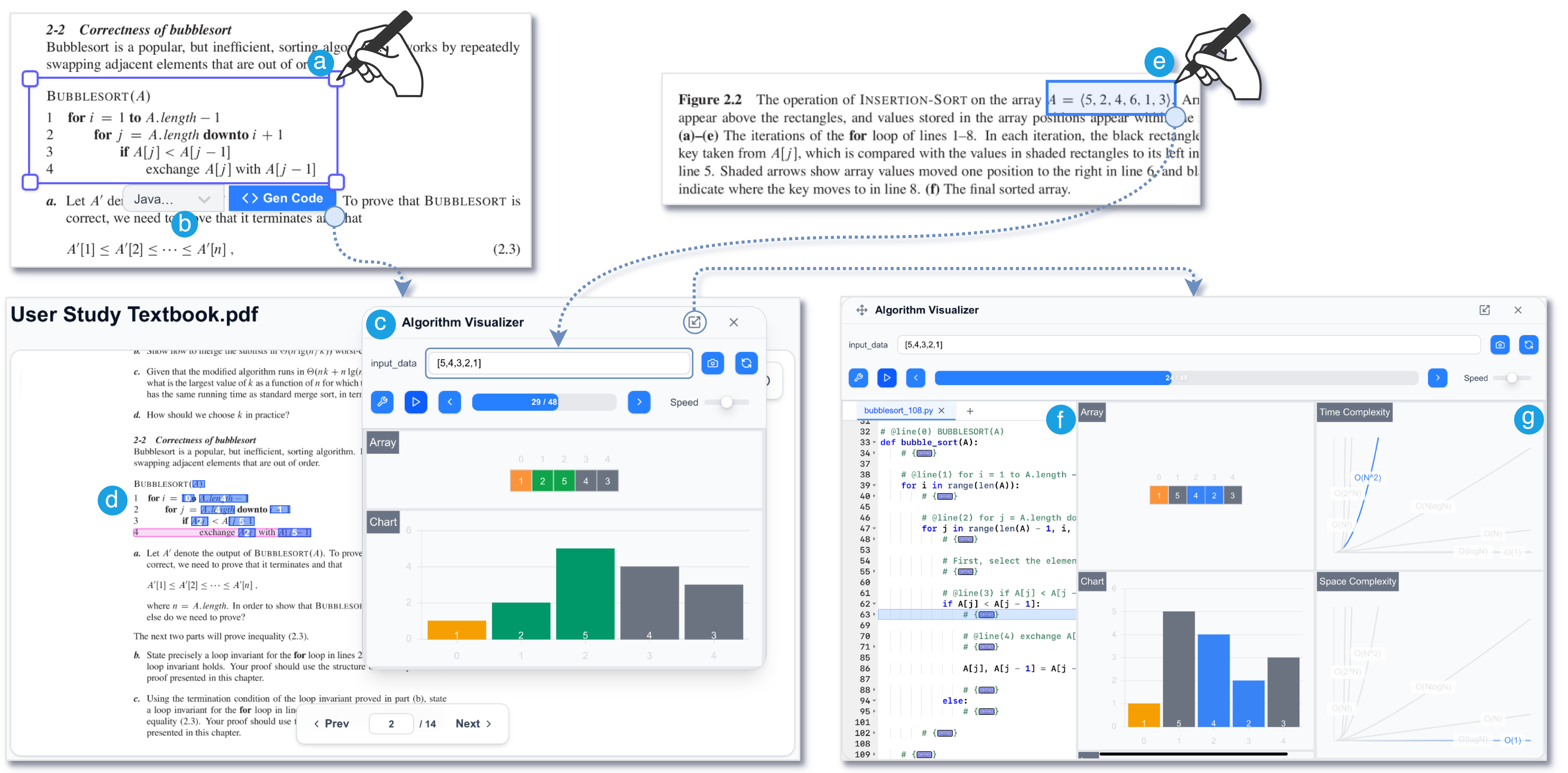}
  \caption{An example of using \sys to transform a static textbook algorithm into executable code and dynamic visualizations. (a)~The user selects pseudocode by drawing a box around it. (b)~A language panel allows them to choose a programming language. (c)~The generated algorithm visualization is displayed in a movable window. (d)~Stepping through the algorithm execution highlights the current line, with variable values shown inline. (e)~Input arrays can be captured directly from the page. (f)~For detailed review and exploration, a code editor supports customization of the algorithm and visualization. (g)~Time and space complexities are shown as a line chart.}
  \label{fig:system-walkthrough}
\end{teaserfigure}


\maketitle

\section{Introduction} \label{sec:introduction}

Data Structures and Algorithms (DSA) are a cornerstone of computer science practice. However, learning and applying algorithmic strategies is not easy because it involves abstract concepts that are often invisible and dynamic in nature~\cite{fouh_role_2012}. As a result, the static presentations of algorithms commonly found in textbooks -- such as textual explanations and static diagrams -- often fall short in conveying the underlying logic and temporal behavior of algorithms. 

To support a better understanding of DSA, researchers have been studying algorithm visualization (AV) \cite{shaffer_algorithm_2010, goel_review_2023}. AV tools graphically represent the concepts of DSA to allow learners to observe their step-by-step execution~\cite{guo_online_2013, shaffer_algorithm_2010}. Prior studies have shown that achieving AV success in education is non-trivial (see \Cref{sec:related_algorithm_visualization}), in part because many prior AV tools have been designed as bespoke artifacts and aligning them with course-specific content  is time-consuming and technically demanding for educators~\cite{naps_exploring_2002, shaffer_getting_2011}. Such tools are also generally \textit{external} and \textit{disconnected}, requiring students to switch between visualizations and course materials, which increases cognitive load and disrupts the continuity~\cite{mayer_splitattention_1998}. As a result, AV tools continue to remain underutilized in real-world educational settings, despite their potential pedagogical benefits such as promoting student engagement and hands-on learning~\cite{isohanni_are_2014, shaffer_getting_2011, fouh_role_2012}.

In this paper, we investigate how to address these challenges by designing and implementing \sys, a novel tool that transforms static pseudocode (e.g., such as from textbooks or lecture slides) into interactive step-by-step animations. By leveraging a multi-stage pipeline based on Large Language Models (LLMs), our system parses pseudocode and generates corresponding executable code in Python and JavaScript. It then simulates algorithm execution on user-definable inputs and produces synchronized animations visualizing the associated data structures. 

While creating AV from static documents is not new (e.g., the \textit{DARA} system~\cite{paredes-velasco_augmented_2023}), we contribute in three key ways related to the design and development of novel user interfaces and user experiences (UIs/UXs) for AV learning.  First, unlike \textit{DARA}, which requires real Java code printed in static documents (with custom labels) and supports only Dijkstra's algorithm, our approach supports several fundamental DSAs directly from printed pseudocode, eliminating the need for real code preparation and supporting breadth.

Second, to motivate the development of our system, we conduct a design strategy and requirements analysis to identify design challenges and goals related to successful UI/UX workflows for AV student learning. Based on these, we inform a backend pipeline and set of UI/UX interaction features tailored for such AV learning. 

Third, we contribute insights from three evaluations on the feasibility and utility of our approach via direct manipulation of static pseudocode: a technical evaluation of the system’s accuracy and output quality, a student-focused usability study (N=15), and expert interviews with DSA instructors (N=6).

Our technical evaluation results demonstrate the system's robustness, achieving over 94\% accuracy in pseudocode extraction and 100\% accuracy in analyzing time and space complexity and code generation, with generated AVs receiving high scores for execution fidelity, data encoding, and visualization type selection. The student-focused study confirms the high usability and engagement of this system compared to traditional, static learning approaches (e.g., textbooks). The expert interviews validate our approach against existing teaching materials and explore its integration into current educational practices. Overall, evaluation feedback suggests that our approach can reduce AV preparation and integration workload and encourage learners' in-depth understanding and engagement. Based on the process of designing, implementing, and evaluating \sys, we reflect on insights and lessons learned and discuss areas for future enhancement as well as the implications that technology-enhanced tools like \sys have for the future of learning and teaching.

\section{Related Work}\label{sec:relatedwork}

\subsection{Algorithm Visualization} 
\label{sec:related_algorithm_visualization}

Traditionally, static materials (e.g., textbooks, lecture slides) are the primary resources used for presenting content in DSA courses. Unfortunately, students often struggle to connect static pseudocode with dynamic execution behavior~\cite{fouh_role_2012}.

\textit{Algorithm Visualization} (AV) represent one possible solution for this, by using graphical representations of data structures and allowing users to execute algorithms step-by-step through interaction and animation~\cite{hundhausen_metastudy_2002, shaffer_algorithm_2010, brown_algorithm_1988}. While early literature reported mixed results on the pedagogical effectiveness of AV ~\cite{hundhausen_metastudy_2002}, subsequent work has demonstrated its potential benefits~\cite{grissom_algorithm_2003, simonak_increasing_2020, urquiza-fuentes_survey_2009}, including improved learning outcomes~\cite{fouh_role_2012, hansen_designing_2002, kehoe_rethinking_2001, schweitzer_interactive_2007} and engagement~\cite{hall_evaluating_2013}. 

Motivated by this, researchers have developed several examples of AV tools (e.g.,~\cite{goel_review_2023, shaffer_algorithm_2010, school_of_computing_national_university_of_singapore_visualgo_2015, noauthor_algorithm_2018, david_galles_data_2011, david_galles_cs_2011, tobias_nothlich_algovisio_2020}). Unfortunately, adopting such tools in practice is non-trivial. Such tools are not easy to integrate into existing pedagogical course materials (e.g., students must switch from an existing resource such as a textbook to an external AV tool, which can increase cognitive load and disrupt the continuity of the learning experience~\cite{mayer_splitattention_1998}), or must use specialized/tailored teaching materials (e.g., interactive AVs can be embedded into e-textbooks, but this restricts the content to only the supported digital resources). Likewise, external tools may have inconsistent notations or implementation details compared to existing curriculum resources (e.g., textbook psuedocode)~\cite{shaffer_getting_2011}. As such, educators often hesitate to use external ``standalone'' or ``resource-restricted'' tools in classroom settings~\cite{naps_exploring_2002, shaffer_getting_2011}.

In contrast to this ``external'' or ``resource restricted'' approach, we contribute a flexible and integrative approach for generating AVs directly from pseudocode found in static materials that already exist in DSA courses. In part, because we employ an LLM and mixed reality approach to read and parse pseudocode, we can accommodate different notations (that trip up existing AV tools) while also supporting existing curriculum materials.

\subsection{Augmenting Static Documents}

Making static resources interactive and explorable is an ongoing goal in technology-enhanced learning. For example, prior work has developed techniques to augment documents with semi-automatically generated summaries~\cite{chen_marvista_2023}, references~\cite{li_holodoc_2019}, highlights~\cite{chi_scenthighlights_2005, fok_scim_2023, kim_facilitating_2018}, annotations~\cite{kang_threddy_2022, hullman_contextifier_2013, subramonyam_texsketch_2020}, and visualizations~\cite{badam_elastic_2019, masson_charagraph_2023, shi_augmenting_2025}. Systems like Augmented Math~\cite{chulpongsatorn_augmented_2023} and Augmented Physics~\cite{gunturu_augmented_2024} extend this line of work by transforming diagrams into interactive, animated media. While we share the goal of making documents interactive, our work significantly differs in the type of content we augment and the inherent challenges: prior research has largely focused on descriptive content (e.g., explaining what a system is), whereas we focus on procedural content that explains how to perform a process, such as an algorithm, or represent the content of a data structure.

The most relevant prior work in this area is DARA~\cite{paredes-velasco_augmented_2023}, a system that uses mobile AR to recognize and visualize Dijkstra's algorithm from a printed page. This work shares our objective of performing AV based on static (printed) materials. However, DARA's approach is limited in generalizability and scalability. As mentioned above, it only works on Dijkstra's algorithm; it also requires pre-placed markers on the printed material, and overlay content must be manually prepared beforehand. In contrast, our system employs a markerless approach that analyzes pseudocode directly from unmodified textbooks on-the-fly and then dynamically generating visualizations. This approach is also more generalizable, by allowing the transformation of any existing DSA textbook into an interactive learning experience across a wide range of common algorithms.

\subsection{Automated Visualizations}
\label{sec:automated_visualization}

Previous research in AV has examined varying degrees of automation in mapping algorithmic procedures to visual representations~\cite{kerren_algorithm_2001}. Early systems~\cite{ben-ari_perspectives_2001, concepcion_algorithma_2000, mornar_system_2014} reduced manual coding effort by either offering semi-automated visualization semantics or introducing specialized pseudocode languages. These approaches tightly coupled to the syntactic structure of algorithmic implementations rather than their semantic significance.

More recently, the emergence of LLMs has supported the flexible and expressive creation of visualizations via natural language. So-called Natural Language to Visualization (NL2VIS) systems~\cite{dibia_lida_2023, vaithilingam_dynavis_2024, tian_chartgpt_2025} generate complete visualization code directly from high-level natural language queries, primarily for tabular data. 

Our work extends this line of research by applying LLM-based visualization generation in the context of AV. Inspired by the notion of \textit{Interesting Events} introduced by BALSA~\cite{brown_system_1984, brown_algorithm_1988}, our pipeline generates executable code with embedded visualization triggers based on parsing pseudocode. Unlike traditional approaches that mechanically map code operations to visual effects, we employ LLMs to semantically analyze algorithmic procedures and strategically insert visualization commands at points judged significant, independent of the specific syntactic operations in the code. This approach enables the generation of more contextually appropriate visualizations that highlight the conceptual essence of algorithms rather than merely reflecting their implementation details.

\section{Design Challenges and Goals} \label{sec:design-challenges-goals}

As discussed in \Cref{sec:introduction,sec:relatedwork}, AV tools are often external and disconnected from traditional course materials, not to mention they have limited flexibility to work with a variety of DSA content notations (e.g., variations in pseudocode or algorithm implmentation details) that are found in CS curricula. To address these issues, we employ a design methodology~\cite{sedlmair2012design} to design, engineer, and evaluate \sys. In alignment with design study practices, we first considered the design challenges and goals of such a potential tool. To distill a set of design challenges, we performed a meta-analysis of existing AV tools based on the literature discussed in \Cref{sec:relatedwork}, 
reviewing their current limitations and how these could be addressed. From this, we derived a set of five salient design challenges \circled{C1}--\circled{C5}: 

\textbf{\circled{C1} Lowering the Barrier for Instructor Adoption:} A major barrier to adopting AVs in classroom is the difficulty of finding or creating suitable animations or presentations that align with course content~\cite{naps_exploring_2002, shaffer_getting_2011}. Instructors need mechanisms that minimize the effort required to prepare effective AVs for their specific curriculum.

\textbf{\circled{C2} Seamless Integration with Existing Learning Materials:} For widespread adoption, AV tools should integrate seamlessly with existing materials like textbooks and lecture notes~\cite{shaffer_getting_2011}, as discrepancies between tool examples and course materials can cause confusion and disrupt learning flow~\cite{naps_exploring_2002}.

\textbf{\circled{C3} Creating Effective and Engaging AVs:} For an AV to be pedagogically effective, it must not only visually represent the algorithm but also actively engage the learner~\cite{hundhausen_metastudy_2002}. Following established best practices for AV design~\cite{naps_exploring_2002}, visualizations should provide elements that promote hands-on engagement and interaction, rather than passive observation.

\textbf{\circled{C4} Diverse Programming Language Backgrounds:} While DSA is language-agnostic, algorithms are often taught in specific programming languages (e.g., Python, Java, C++). Tools fixed to an unfamiliar language impose a dual cognitive load of learning both the algorithm's logic and language syntax. Therefore, learners need a flexible environment that lets them focus on the algorithm using a language they are already proficient in.

\textbf{\circled{C5} Preserving Learner Agency:} While generated code and visualizations are convenient, they should not be unmodifiable ``black boxes.'' Related to \circled{C3}, there is strong evidence for the benefits of experiential learning through active participation~\cite{hundhausen_metastudy_2002, hudson_designing_2008}, such as students modifying a data structure and observing cause and effect on the algorithm's results and runtime. The challenge is to balance the convenience of automation while ensuring learners retain agency to edit and experiment with the code.

Then, to address the design challenges \circled{C1}--\circled{C5}, we derived a set of five corresponding design goals \circledVarD{G1}--\circledVarD{G5}:

\textbf{\circledVarD{G1} Automated Algorithm Visualization Generation:} To lower the barrier for instructor adoption (\circled{C1}), we aim to automatically generate AVs directly from static pseudocode, eliminating the need for instructors to manually create or search for suitable animations, thus streamlining the integration of AVs into their curriculum.

\textbf{\circledVarD{G2} Tightly-Coupled Learning Material Integration:} To ensure seamless integration with existing curriculum materials (\circled{C2}), we aim to overlay AVs directly onto existing materials such as static textbooks or lecture slides, coupled with UI/UX features such as interactively animating the AV execution paired with code highlighting and variable tracing to bridge the gap between abstract pseudocode and its concrete execution.

\textbf{\circledVarD{G3} Pedagogically Sound Visualizations:} To create effective and engaging learning experiences (\circled{C3}), we want AVs to be guided by established visualization design practices~\cite{naps_exploring_2002}, focusing on principles such as ``provide multiple views,'' ``support flexible execution control,'' and ``complement visualizations with explanations,'' to encourage active exploration rather than passive observation.

\textbf{\circledVarD{G4} Multi-Language Support:} To accommodate learners from diverse programming backgrounds and reduce cognitive load from unfamiliar syntax (\circled{C4}), we aim to support multiple programming languages by decoupling the executable algorithm code from its visual representation, allowing students to engage with the material in a familiar programming environment.

\textbf{\circledVarD{G5} Flexible Editing of Algorithm Visualization:} To preserve learner agency (\circled{C5}) and foster deeper engagement (\circled{C3}), we intend to provide flexible editing capabilities, such as allowing users to directly modify the algorithm, the visualization code, and the input datasets. This empowers them to correct auto-generation errors, explore ``what-if'' scenarios, and even construct their own visualizations, an activity known to enhance learning~\cite{stasko_using_1997}.

\section{System Design and Implementation}

\sys is a web-based, LLM-integrated AV tool designed to fulfill \circledVarD{G1}--\circledVarD{G5}. It supports DSA learning by transforming static pseudocode into interactive, step-by-step AVs, allowing users to explore algorithm execution, performance metrics, and data structure transformations in real time. Below, we describe the system architecture and interface design details of \sys; evaluations are presented in Sections~\ref{sec:tech_eval} and~\ref{sec:user_study}. 
To support reproducibility, the system's codebase is available at \textsf{<GitHub link anonymized for review round>}.


\subsection{Learning Workflow}


To help scope the workflows enabled by our system, consider the DSA topics that are commonly taught in CS curriculums, such as sorting algorithms, linked list data structures, and binary search tree operations. An example learning workflow for these using \sys can comprise the following steps: 1)~capture (or upload) an image of a curriculum material (e.g., a textbook page, lecture slide, PDF handout, etc.), 2)~highlight the pseudocode and select the desired programming language, 3) generate and execute an AV instance, and 4) support follow-up learning and exploration via interactions with the AV. Below, we illustrate this workflow using a bubble sort example from \textit{Introduction to Algorithms}~\cite{cormen2009introduction}, one of the most popular university-level introductory DSA textbooks in the United States. Figure~\ref{fig:system-walkthrough} illustrates this walkthrough.

\textbf{Step 1: Capture or Import a DSA Page:}
A user first captures a snapshot of a static DSA resource (e.g., a textbook page) and imports it into \sys either via a camera on the user's laptop/tablet or by importing a PDF or other image file format. 
The snapshot is converted to a standardized image format for processing. The system performs layout detection using OCR and computer vision to extract pseudocode, and overlays bounding boxes on detected pseudocode regions to guide user selection in the subsequent step (\circledVarD{G2}).

\textbf{Step 2: Select Pseudocode and Programming Language:}
The user next selects the pseudocode by adjusting a bounding box using corner handles to encompass the entire code block (\Cref{fig:system-walkthrough}-a) and chooses a programming language for code generation from two supported options: Python or JavaScript (\circledVarD{G4}) (\Cref{fig:system-walkthrough}-b).

\textbf{Step 3: Generate and Execute an Algorithm Visualization Instance:}
The system converts selected pseudocode into executable code for the selected programming language, injected with additional AV commands (\circledVarD{G1}). Specifically, the generated code incorporates hook functions that visualize dynamic representation of algorithmic operations and data structure modifications.
The code is processed through a visualization engine to produce JSON-formatted commands defining animation sequences, timing, and visual effects, which are rendered into an interactive algorithmic demonstration to the user (\Cref{fig:system-walkthrough}-c), along with visualizations of worst-case time and space complexity alongside the animation (\circledVarD{G3}) (\Cref{fig:system-walkthrough}-g).

\textbf{Step 4: Supporting Interactions with the Algorithm Visualization:}
Several features support further interaction and editing.
Animation can be triggered to observe additional synchronized overlays on the original pseudocode (i.e., on the textbook), including code-line highlighting and variable tracing, which supports step-by-step exploration of algorithmic execution (\circledVarD{G2}, \circledVarD{G3}) (\Cref{fig:system-walkthrough}-d).
The AV can be played with default randomized datasets or user-provided inputs. A capture function lets users to extract numerical values or arrays directly from DSA resources (e.g., other textbook pages) via OCR and use them as input data (\Cref{fig:system-walkthrough}-e), and an integrated code editor allows further modification of the algorithm and visual representation (\circledVarD{G3}, \circledVarD{G5}), if desired (\Cref{fig:system-walkthrough}-f).

\subsection{System Architecture and Pipeline}
\begin{figure*}[t]
    \includegraphics[width=.9\linewidth]{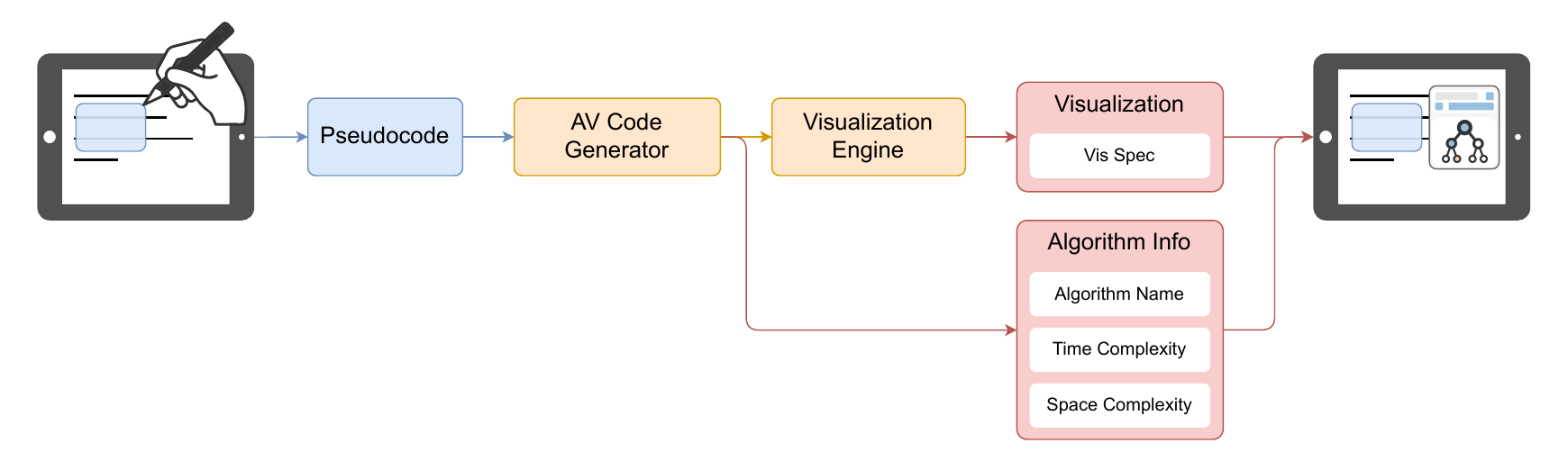}
    \vspace{-1em}
    \caption{An overview of the \sys system architecture.}
    \label{fig:system-architecture}
\end{figure*}

The system comprises three main components: frontend, backend, and visualization engine. The frontend is implemented using React.js and D3~\cite{bostock_d3_2011} for dynamic visualization rendering. 
The backend is implemented using FastAPI\footnote{https://fastapi.tiangolo.com} to orchestrate core system functionalities, including OCR processing for textbook analysis and LLM integration for automatic code generation. For textbook analysis, we employ Google Cloud Vision API\footnote{https://cloud.google.com/vision/docs/ocr} for text extraction and LayoutParser~\cite{shen_layoutparser_2021} for detecting pseudocode regions. For code generation, we utilize LangGraph\footnote{https://www.langchain.com/langgraph} to implement the AV Code Generator, coordinating multiple LLM-based agents to transform pseudocode into executable algorithms with integrated visualization hooks.
The visualization engine operates as a serverless AWS Lambda\footnote{https://aws.amazon.com/lambda/}-based execution environment supporting multiple programming languages. This component executes the generated algorithm code and produces JSON-formatted visualization commands. An overview of the main system architecture is shown in \Cref{fig:system-architecture}; we detail the steps below.

\textbf{Pseudocode Extraction:}
The system first processes the captured content (e.g., a PDF or image input) by first converting it into a standard image format for layout analysis. It then applies LayoutParser~\cite{shen_layoutparser_2021} to segment documents into structural elements such as \textit{text}, \textit{title}, \textit{list}, \textit{table}, and \textit{figure}. The system filters for \textit{list}, \textit{table}, or \textit{figure} elements, as these frequently contain pseudocode. The bounding box coordinates of these candidate regions are returned to the frontend, which overlays rectangular highlights on the textbook to assist users selection. Once a region is selected, the system captures the pseudocode within that region and converts it to textual data using Google Cloud Vision API. This OCR processing transforms the visual pseudocode representation into text format, enabling the subsequent analysis and code generation processes.

\begin{figure}[htbp]
    \includegraphics[width=.6\linewidth]{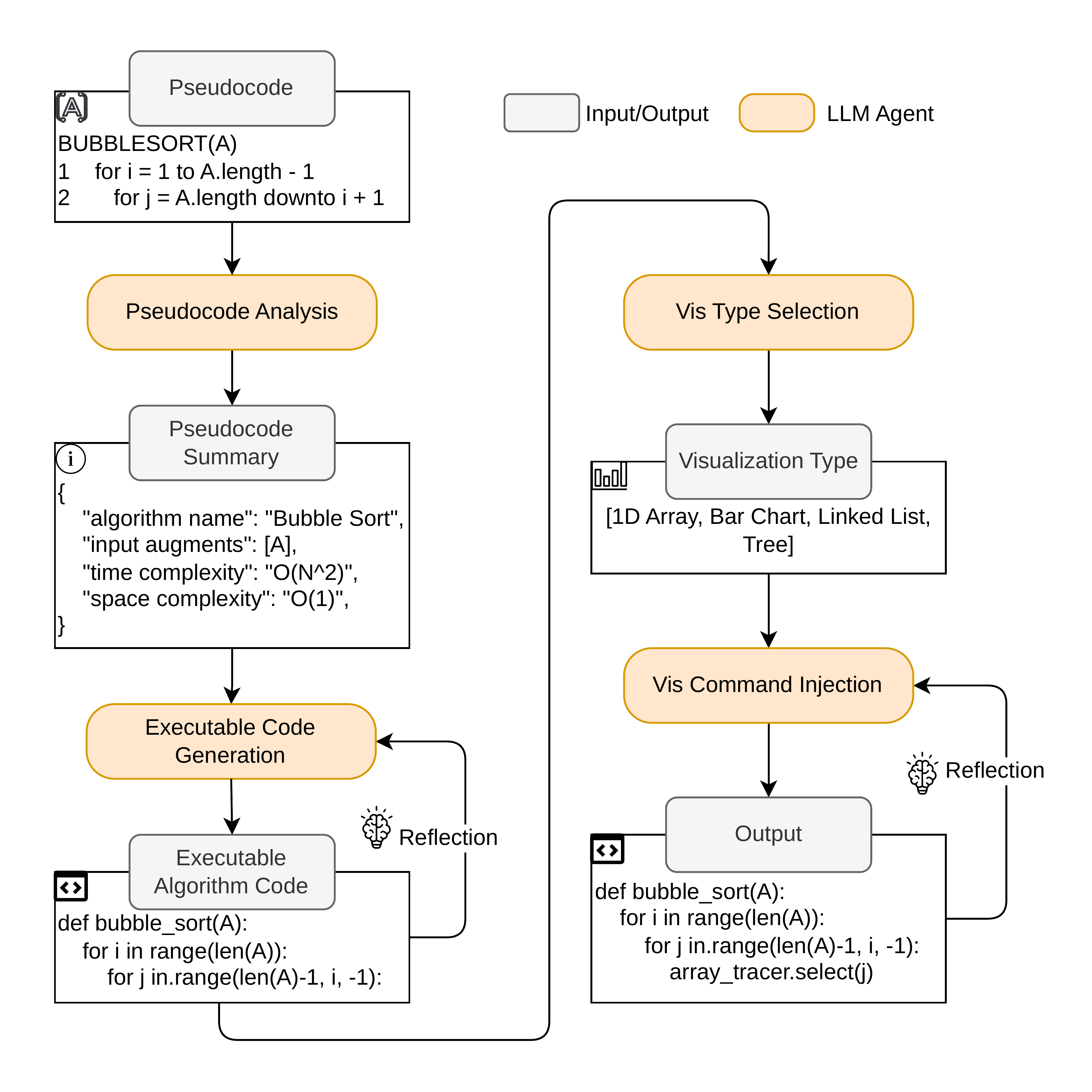}
    \caption{The AV Code Generator pipeline for generating executable visualization code.}
    \label{fig:llm-pipeline}
\end{figure}

\textbf{AV Code Generation:}
The AV Code Generator component receives the pseudocode text and then initiates the process to generate AV code. Specifically, this component comprises four LLM-powered modules: \textit{pseudocode analysis}, \textit{executable code generation}, \textit{visualization type selection}, and \textit{visualization command injection}, orchestrated using LangGraph (\Cref{fig:llm-pipeline}). We strategically employ different LLM models based on task complexity: GPT-5 nano\footnote{https://platform.openai.com/docs/models/gpt-5-nano} for lightweight analysis tasks (\textit{pseudocode analysis} and \textit{visualization type selection}), and Claude Sonnet 4\footnote{https://www.anthropic.com/news/claude-4} for complex code generation tasks (\textit{executable code generation} and \textit{visualization command injection}).

The \textit{pseudocode analysis} module analyzes the textual pseudocode to infer 1) the algorithm name, 2) required input parameters, and 3) time/space complexity.
The \textit{executable code generation} module produces runnable code in the user's selected programming language that faithfully implements the analyzed pseudocode. To ensure code reliability, this module incorporates a reflection mechanism that automatically detects and corrects execution errors through iterative debugging.
These outputs are then passed to the \textit{visualization type selection} module, which determines the most appropriate visualization type (e.g., one-dimensional arrays, bar charts, linked lists, and tree node-link diagrams, see \Cref{fig:frame-by-frame-animation}) based on the algorithm's data manipulation patterns.
Finally, the \textit{visualization command injection} module (which also includes a reflection mechanism) inserts visualization functions at points where interesting events occur. 

\textbf{Visualization Command Extraction and Serialization:} 
The Visualization Engine executes the generated code within a Lambda environment corresponding to the specified programming language, capturing visualization commands during algorithm execution. Each visualization operation is intercepted and converted into structured JSON commands that encode the operation type, target elements, and parameters. This standardized JSON format is language-agnostic, enabling the frontend to process visualization commands uniformly regardless of the underlying programming language. Each command contains three components: a key for visualization instance identification, a method name for the visualization operation, and an arguments array containing method parameters. This format enables deterministic reconstruction of the visualization sequence in the frontend.

\textbf{AV Rendering:}
Finally, the frontend receives the JSON command sequence and processes it through a command interpreter that maps operations to data manipulation functions. The system maintains internal data structures bound to visualizations through D3 data joins.
Command execution updates the internal data model, triggering D3's data-driven rendering to generate animations. Each command corresponds to specific visual changes, maintaining synchronization between algorithm execution and display.
The fontend panel provides a set of interactive playback controls, allowing users to navigate the command sequence at variable speeds,
pause at algorithmic steps, or advance frame by frame.

\subsection{AV Interaction Features}

To support our learning workflow, we design and integrate a set of interactive functionalities in \sys, including \textit{interactive frame-by-frame animation}, \textit{dynamic input}, \textit{synchronized code-line highlighting}, \textit{variable tracing}, and \textit{time/space complexity visualization}. Together, these features transform static pseudocode into a dynamic and exploratory environment, enabling learners to observe algorithm behavior in real time and engage directly with both code and data.

\begin{figure}[h]
  \includegraphics[width=.6\linewidth]{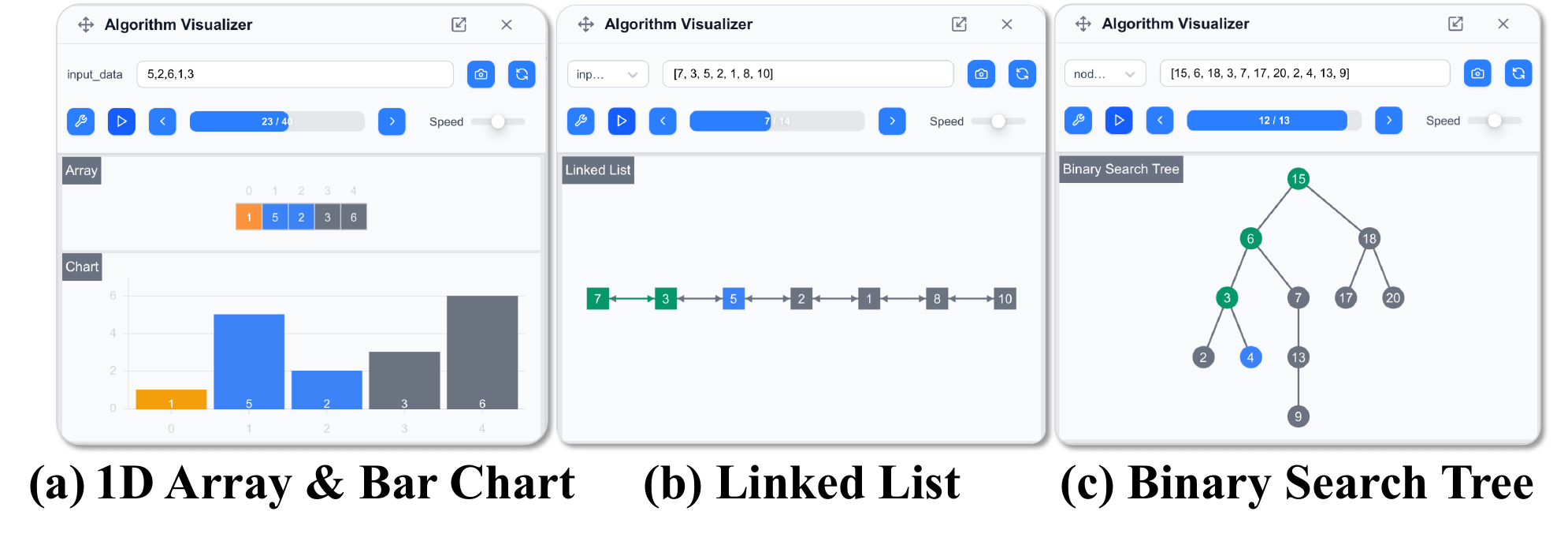}
  \caption{Visualization windows showing frame-by-frame animation controls.}
  \label{fig:frame-by-frame-animation}
\end{figure}

\textbf{Interactive Frame-by-Frame Animation:}
Animation has been shown to be an effective mechanism for explanatory visualization~\cite{lin2010effect}, hence we employ it here.
This feature allows learners to follow and control an algorithm’s execution step by step, supporting agency in observing both the algorithm progression and changes in data structures over time via Play, Next, and Previous controls.
\Cref{fig:frame-by-frame-animation} shows examples of how the system supports three common data structures: 1)~one-dimensional arrays, 2)~linked lists, and 3)~trees. (The system is also extensible to support additional data structures and algorithms; see \Cref{sec:discussion}.) 
Arrays are displayed in a grid layout with bar charts for numerical data; linked lists show nodes connected by directional arrows with pointer indicators; and trees are rendered using hierarchical layouts.
During animation, each visualization supports dynamic operations such as value modification, element highlighting, and node/edge insertion or deletion. 

\begin{figure}[h]
  \includegraphics[width=.6\linewidth]{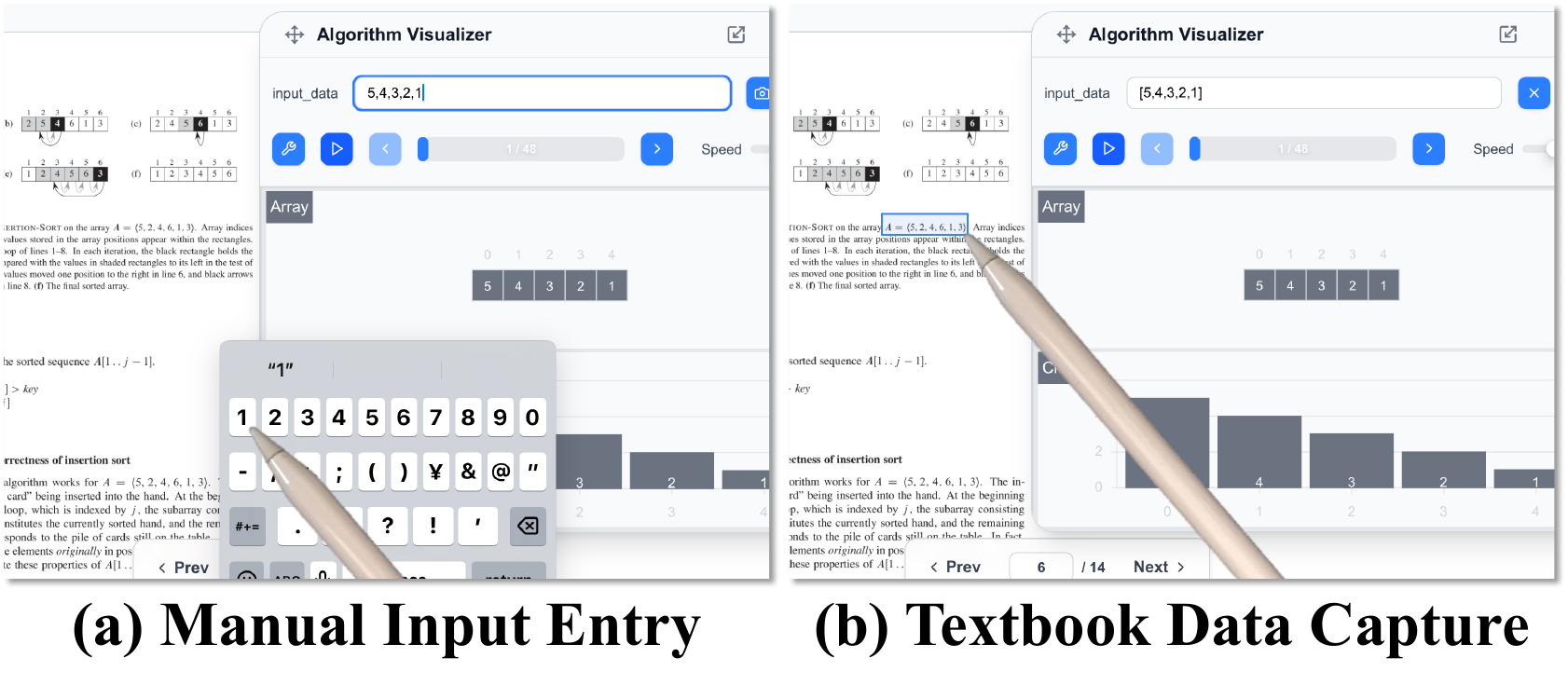}
  \caption{An example of dynamically inputting data structure values.}
  \label{fig:dynamic-input}
\end{figure}

\textbf{Dynamic Input:}
Users can provide custom input values through manual entry via input forms or direct capture from textbook examples using the integrated capture function. This dual approach allows learners to test textbook scenarios or any values they wish to explore, providing a scaffold for experiential learning about algorithm behavior under different conditions, and support tighly-coupled integration that leads to the automatic generation (or updating) or AVs. The system by default supports three input types: single numerical values, one-dimensional arrays, and two-dimensional arrays, though, similar to the data structured mentioned above, the system is extensible for additional input types.
\Cref{fig:dynamic-input} shows an example of how users can update an input array by entering values in the form or capturing data directly from a textbook example.

\begin{figure}[h]
  \includegraphics[width=.6\linewidth]{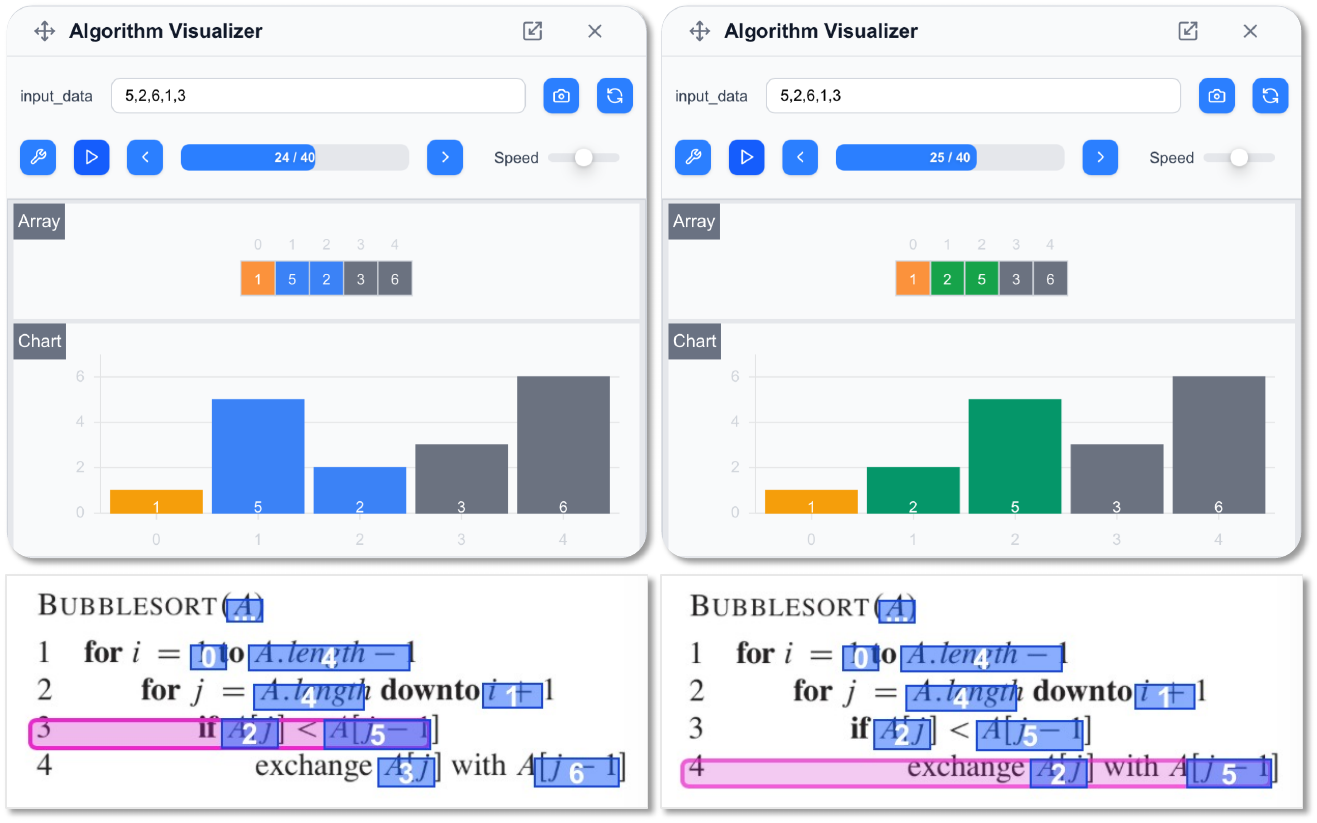}
  \caption{Synchronized code-line highlighting and variable tracing.
  }
  \label{fig:tracing-and-highlighter}
\end{figure}

\textbf{Synchronized Code-Line Highlighting:}
During animation, the currently executing line of pseudocode is dynamically highlighted on the resource image (i.e., the textbook page). This real-time synchronization enables learners to see exactly how each line of code translates into algorithmic actions, with the goal of making abstract algorithmic concepts immediately apparent in terms of code execution.
For example, \Cref{fig:tracing-and-highlighter} illustrates this feature. The executed lines (Lines 3 and 4) are indicated with a semi-transparent pink rectangle, and the highlight dynamically shifts as the animation advances through each step of the algorithm.

\textbf{Variable Tracing:}
Variable values are overlaid on the pseudocode and dynamically update as the algorithm executes. The goal is to provide learners with an intuitive understanding of how data evolves and flows at each step. 
For example, \Cref{fig:tracing-and-highlighter} illustrates variable tracing, where each variable is annotated with its current value inside a semi-transparent blue rectangle. With these overlaid values, learners can clearly see that at Line 3, the condition $A[2] (=2) < A[1] (=5)$ evaluates to true, and at Line 4, the values of $A[2] (=2)$ and $A[1] (=5)$ are exchanged.

\begin{figure}[h]
  \includegraphics[width=.4\linewidth]{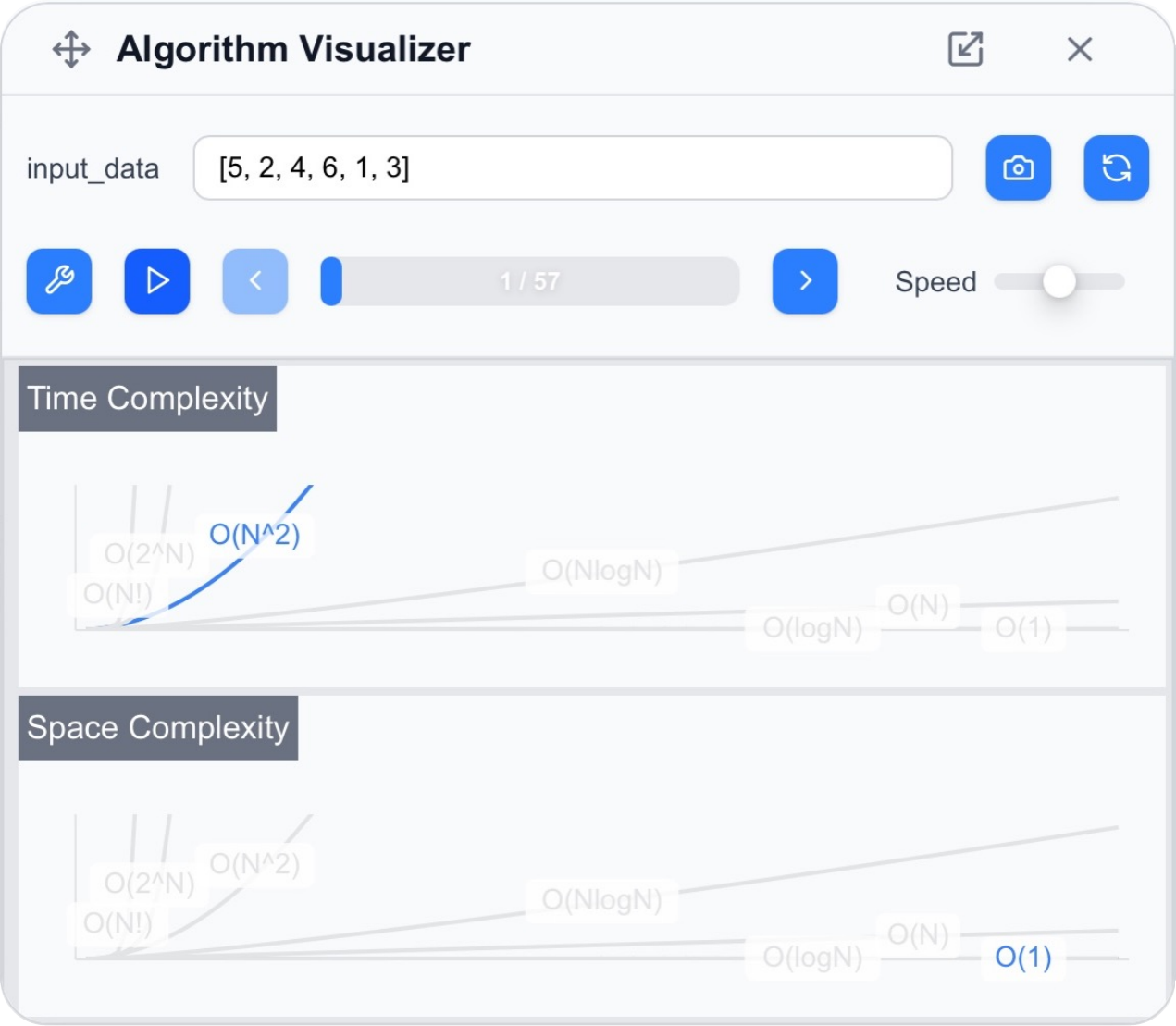}
  \caption{The time/space complexity visualization window.}
  \label{fig:time-space-complexity}
\end{figure}

\textbf{Time/Space Complexity Visualization:}
Finally, the system presents theoretical performance metrics, including time and space complexity in Big-O notation, with a static visualization of worst-case scenarios to help learners understand the theoretical upper bounds of algorithm performance and resource usage. For example, \Cref{fig:time-space-complexity} illustrates the time and space complexity of bubble sort.

\section{Technical Evaluation}
\label{sec:tech_eval}

To evaluate the computational accuracy and quality of our backend pipeline, we collected 23 DSA textbooks and randomly selected 10 pseudocode examples each for insertion sort, linked-list traversal, and binary search tree search. We applied our pipeline to each sample and evaluated the correctness and quality of the generated animations using a hybrid automatic and manual assessment.

For accuracy evaluation, we considered three aspects. First, we measured the character error rate (CER) for pseudocode OCR detection. Specifically, we computed the Levenshtein distance between the OCR output and the ground-truth pseudocode and defined accuracy as:
\begin{displaymath}
\text{accuracy} = 1 - \frac{\text{Levenshtein distance}}{\text{ground-truth length}}
\end{displaymath}
Second, we assessed the accuracy of time and space complexity analysis by comparing LLM predictions against ground-truth complexity from the textbooks, calculating the proportion of correctly matched results. Third, we evaluated the correctness of the generated executable algorithm code (i.e., prior to the injection of visualization commands) using unit tests and manual verification. Due to the wide syntactic variation in pseudocode (e.g., Python-like syntax, natural-language-based syntax), one of the authors, who had completed an advanced data structures and algorithms course, manually verified that generated code preserved the structure and semantics of the original pseudocode, while unit tests ensured correct execution.

Next, to assess the quality of generated AVs, we designed an animation rubric encompassing \textit{Execution Fidelity}, \textit{Data Encoding}, \textit{Visualization Type}, and \textit{Aesthetics} (\Cref{table:av_quality_rubric}), as no standardized method exists for evaluating AVs. To minimize individual bias, two authors independently rated each animation using a 5-point scale (1=poor, 5=excellent), and the final score was computed as the average of their ratings.

\Cref{tab:tech-evaluation-results} summarizes the results of these evaluations, which are described in detail below. 

\begin{table}[h]
    \centering
    \caption{Evaluation rubric for assessing AV quality.}
    \label{table:av_quality_rubric}
    \begin{tabularx}{0.48\textwidth}{lX}
        \toprule
        \textbf{Dimension} & \textbf{Question} \\
        \midrule
        \textbf{Execution Fidelity} & Are all algorithmic steps executed correctly and in proper sequence, without omissions or hallucinations? \\
        \hline
        \textbf{Data Encoding} & Does the visualization accurately reflect the data structure state at every algorithmic step? \\
        \hline
        \textbf{Visualization Type} & Considering the algorithm's features, is the visualization type appropriate for representing the algorithmic process? \\
        \hline
        \textbf{Aesthetics} & Are the aesthetics of the visualization elements (color, shape, position) appropriate and effective for the algorithm? \\
        \bottomrule
    \end{tabularx}
\end{table}

\subsubsection*{\textbf{Accuracy Results}}
For pseudocode OCR detection, our pipeline achieved 94.1\% for insertion sort, 95.4\% for linked list traversal, and 94.7\% for binary search tree search. For time and space complexity analysis and executable code generation, the system achieved 100\% accuracy. Notably, the LLM correctly inferred complexity even when the same algorithm employed different processing strategies (e.g., recursion vs. loop iteration). These results demonstrate the robustness of the LLM's inference capabilities, accurately inferring complexity and generating executable code even when OCR results were incomplete.

\subsubsection*{\textbf{AV Quality Results}}
The quality evaluation revealed varying performance across different criteria. All algorithms achieved perfect scores (5.0) for \textit{Execution Fidelity}, indicating that visualization command injection did not disrupt the generated code.
\textit{Data Encoding} scores were consistently high: insertion sort and binary search tree achieved 5.0, while linked list search scored 4.6. The lower score for linked-list search was attributed to a mismatch between the pseudocode implementation and visualization requirements, where the pseudocode contained only a partial linked list class implementation without essential initialization functions, which led to failure in creating the linked list.
\textit{Visualization Type} achieved perfect scores (5.0) across all algorithms. \textit{Aesthetics} presented the lowest scores (insertion sort: 3.7, linked list: 2.9, binary search tree: 3.3). While the system successfully applied color coding to elements during active operations, visual feedback for completion states and traversal indicators sometimes failed to meet expectations and showed inconsistency within the same algorithm type. Specifically, color coding to indicate sorted completion or traversal history was sometimes not implemented as expected and exhibited substantial variation even among the same algorithm samples. Despite the rather underwhelming \textit{Aesthetics} score in this evaluation, the follow up usability evaluations actually still found quite high overall usability, as discussed in the next section (see \Cref{sec:discussion} for further discussion).

\begin{table*}[t]
    \centering
    \caption{Evaluation results for three algorithms. 
    LLM pipeline accuracy is reported as percentages, and AV quality is rated on a 5-point scale (1=poor, 5=excellent).}
    \label{tab:tech-evaluation-results}
    \footnotesize{
    \begin{tabular}{lccccccc}
    
    \toprule
    & \multicolumn{3}{c}{\textbf{Accuracy}} & \multicolumn{4}{c}{\textbf{AV Quality (5-point scale)}} \\
    \cmidrule(lr){2-4} \cmidrule(lr){5-8}
    \textbf{Algorithm} & 
    \makecell{Pseudocode \\ Extraction} & 
    \makecell{Time/Space \\ Complexity} & 
    \makecell{Executable Code \\ Generation} & 
    \makecell{Execution \\ Fidelity} & 
    \makecell{Data \\ Encoding} & 
    \makecell{Visualization \\ Type} & 
    Aesthetics \\
    \midrule
    Insertion Sort        & 94.1\% & 100\% & 100\% & 5 & 5 & 5 & 3.7 \\
    Linked List    & 94.7\% & 100\% & 100\% & 5 & 4.6 & 5 & 2.9 \\
    Binary Search Tree    & 95.4\% & 100\% & 100\% & 5 & 5 & 5 & 3.3 \\
    \midrule
    \textbf{Average}      & \textbf{94.7\%} & \textbf{100\%} & \textbf{100\%} & \textbf{5} & \textbf{4.87} & \textbf{5} & \textbf{3.28} \\
    \bottomrule
    \end{tabular}
    }
\end{table*}

\section{Empirical Evaluations}
\label{sec:user_study}
Next, to evaluate the usability aspects and pedagogical implications of our system, we conducted a pair of user studies: (1) a usability study with 15 students, and (2) expert interviews with 6 university-level computer science educators. The usability study focused on assessing ease of use and feature preferences, while the expert interviews collected insights into the system's educational potential from professional educators' perspectives.

\subsection{Usability Study}
\label{sec:usability_study}
We evaluated the usability of our system via its tablet interface with pen-based interactions. To ensure comparability, identical pages from \textit{Introduction to Algorithms}~\cite{cormen2009introduction} were employed across all conditions. A total of 15 undergraduate and graduate computer science students (5 males, 10 females, ages 20--25 (m=22.2)) from \textsf{Anonymized University} participated in the study (see \Cref{appdix:demographics} for demographics). All participants had previously completed an introductory DSA course. This was a crucial requirement, as we did not want participants to have to freshly learn the underlying concepts immediately prior to taking the study, and could thus focus on evaluating system usability aspects and feature preferences. Moreover, this also mimics a likely real-world use case for \sys: participants would first be exposed to a DSA topic via ``traditional'' means (e.g., via textbook or lecture), and then would reinforce concepts via reviewing such content via \sys. Study sessions were conducted in person and lasted approximately 35 minutes each. 

\textbf{Method: }
To ensure consistent conditions across all participants in the user study, we used pre-generated AVs whose correctness was verified (see \Cref{sec:discussion} for further discussion).

After obtaining user consent, participants were introduced to the system through a tutorial video and guided walk-through of its features. Once they had become familiar with the interface, they were asked to complete four tasks for each of three algorithms (i.e., insertion sort, linked list search, binary search tree search) to evaluate different functionalities of the system.
At the beginning of each task, participants first read the pseudocode and surrounding explanation directly from the textbook, without using any interactive features. This served as the baseline condition, allowing participants to experience a ``traditional'' textbook-based learning process immediately before interacting with \sys. After completing this reading phase, participants used the tool's interactive features to answer the following questions:

\begin{itemize} 
    \item What are the worst-case time and space complexity?
    \item When a specific element is found (or placed), which line of code is executed?
    \item What is the value of a key variable (e.g., loop index or function argument) at that moment?
    \item When the algorithm fails to find the target, which line is executed and what value is returned?
\end{itemize}

After completing all tasks, participants filled out a post-study questionnaire. The survey gathered feature-specific feedback and overall engagement and usefulness of the system. Feature-specific usefulness was evaluated with statements such as ``Frame-by-frame animation was useful,'' and overall engagement and usefulness with statements such as ``I felt studying with the textbook with this system was useful,'' both rated on a 7-point Likert scale (1=strongly disagree, 7=strongly agree). Additional open-ended questions elicited qualitative feedback. Finally, participants completed the System Usability Scale (SUS)~\cite{john_sus_1995} for the system.

\textbf{Results: }
\sys received a high SUS score of 92.17 (SD=5.73), and participants felt they could use the system quickly and successfully (see \Cref{appdix:sus} for summary of SUS result).

\textbf{\textit{Comparison of Textbooks and Augmented Algorithm}: }
The results indicated substantial differences between the two conditions. Participants rated \sys significantly higher for engagement (M=6.67, SD=0.49) than simply reviewing the textbook directly (M=3.8, SD=1.26) and also for usefulness (M=6.8, SD=0.41) over textbooks (M=3.73, SD=1.28).

Qualitative feedback provided deeper insights into these quantitative differences. For engagement, participants consistently contrasted \sys with passive textbook reading. P11 mentioned that \textit{``seeing the live demo along with which part of the code is being executed made me feel very involved.''} 
This highlights how real-time visualization and code alignment increased participants’ sense of involvement. Similarly, P3 noted that \textit{``interacting with different system components made me feel involved,''} emphasizing the role of active manipulation in fostering engagement.

For usefulness, participants frequently emphasized how visualization clarified complex concepts. P3 remarked, \textit{``Being able to visualize the algorithm in different ways helped me process the information better than just reading the textbook.''} Several participants also highlighted efficiency benefits. P9 noted, \textit{``It is faster to understand through the system compared to studying from the textbook, and visually I can remember things for a longer time.''} These comments suggest that the system not only improved conceptual clarity but also reduced learning time and supported retention.

\begin{figure}[t]
    \centerline{\includegraphics[width=.6\linewidth]{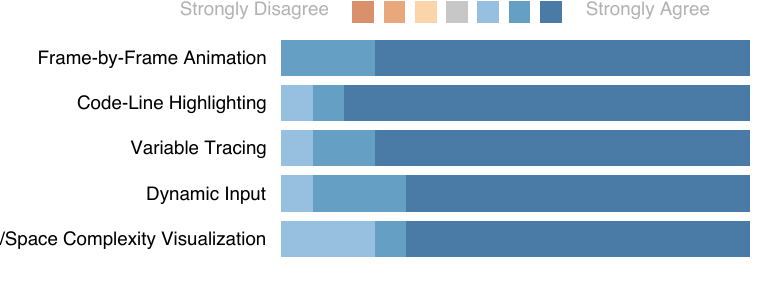}}
    \caption{Usefulness ratings for various system features.}
    \label{fig:feature-comparison-result}
\end{figure}

\textbf{\textit{Comparison of the Different Features}: }
Participants provided generally positive feedback for all features on a 7-point Likert scale (see \Cref{fig:feature-comparison-result}). 
The \textit{frame-by-frame animation} received a high average rating (M=6.80, SD=0.41), with participants appreciating its support for visually tracking the algorithmic process. 
However, P4 noted, \textit{“When the animations are playing and the graphs are changing, it was sometimes hard to understand why that was happening.”}
The \textit{code-line highlighting} feature was similarly rated (M=6.80, SD=0.56). 
Participants valued its ability to indicate the currently executed line of code. 
Several participants further emphasized that combining \textit{frame-by-frame animation} with \textit{code-line highlighting} facilitated comparison between the pseudocode and its visual representation.
Most participants also found \textit{variable tracing} helpful (M=6.73, SD=0.59). 
P6 and P14 mentioned that integrating \textit{code-line highlighting} with \textit{variable tracing} helped them understand each step of the pseudocode. 
P1 commented, \textit{“The variable tracing and code-line highlighting were especially useful since they made dry runs much easier.”} 
However, P8 suggested, \textit{“It would be more convenient to keep this in a separate window and display all variable names with their corresponding values at each step, similar to a debugger.”}
The \textit{dynamic input} feature was also positively received (M=6.67, SD=0.62), as it allowed participants to observe how the algorithm behaved with different inputs. 
Finally, participants appreciated the \textit{time/space complexity visualization} (M=6.50, SD=0.83), which helped them understand algorithmic performance without closely examining the code. 
P6 remarked, \textit{“Having a complexity visualizer helped me understand how some functions are more efficient than others.”} 
Nevertheless, P2 suggested that \textit{“there could be additional animations to explain time and space complexity.”}

\subsection{Expert Interviews}
For our second empirical evaluation, we conducted semi-structured interviews with six instructors (E1--E6) to evaluate the system's potential for educational applications.
All participants had more than 10 years of teaching experience in university-level classes, with five instructors having specific DSA teaching experience (2-10+ years) and one instructor specializing in computer programming (10+ years). Each interview lasted approximately 40 minutes.


\textbf{Method: }
We conducted semi-structured expert interviews through either on-site sessions or online sessions via Zoom. The purpose was to gather feedback on our system, assess its potential educational value, and identify areas for improvement. Each interview followed a common procedure while allowing participants to elaborate on topics based on their own teaching experience. First, we asked instructors to describe their current teaching practices, including the educational materials and tools they use, as well as challenges and pain points they encounter when teaching DSA or programming concepts. We then demonstrated the system and its main functionality. Following the demonstration, participants were given an opportunity to interact with the system themselves, using iPads for on-site sessions or personal computers for online sessions. As part of this hands-on exploration, participants generated algorithm visualizations (AVs) using examples drawn from the same dataset used in our technical evaluation.

After interacting with the system, we asked open-ended questions about their experience and its potential use in educational settings. The questions focused on five aspects: (1) educational value, (2) comparison with existing tools and resources, (3) classroom applicability, (4) support for students' self-study and engagement, and (5) missing pedagogical features. Follow-up questions were asked based on participants' responses to clarify their feedback or explore examples from their teaching practices in greater depth. We organized the resulting interview feedback based on recurring observations across participants and grouped it into the four themes reported below.

\textbf{Results: }
Experts generally felt that our system could assist them in creating AVs for their students, and discussed a number of pedagogical implications. This feedback is summarized below, organized by key findings:

\textbf{\textit{Reducing Instructor Workload and Improving Student Access}: }
Instructors generally relied on lecture slides, blackboard demonstrations, or handwritten notes captured with document scanners to illustrate algorithmic steps. Most experts noted the time-consuming nature of searching for or creating animations independently. Although most were aware of existing online AV systems, they tended to avoid proactively integrating them into their classes. E2 mentioned that \textit{``I usually advise students not to use internet-available AVs because they often employ pseudocode or code that differs from what we use in lecture.''} This highlights the difficulty of aligning external AVs with course-specific materials. 
Our system has the potential to reduce instructors' workload by automatically generating animations that correspond directly to their own pseudocode and implementations. From a student perspective, this alignment also simplifies access to relevant learning resources. As E5 noted, \textit{``Unlike choosing from a long list of algorithms or watching hundreds of videos, students can directly select the content that is most relevant to their own learning needs.''} 
To further support classroom integration, E4 suggested adding video export functionality, stating, \textit{``It would be helpful to export both the animations and highlighted pseudocode so I can include them directly in lectures.''}

\textbf{\textit{Encouraging In-Depth Self-Study}: }
Experts identified several features that support independent learning. 
The clear correspondence between pseudocode and animation was viewed as especially helpful for beginners (E3, E4, E5, E6), as it clarifies abstract algorithmic steps. 
As E4 noted, \textit{``Seeing the pseudocode and the animation side by side really helps students understand each step, especially those who are new to algorithms.''}
Educators also appreciated the system’s ability to link pseudocode with executable code, supporting students’ implementation skills. 
Although the current system supports only a subset of DSA algorithms (see Section~\ref{sec:discussion}), its potential to generate AVs from any pseudocode was highly valued, particularly for specialized or advanced algorithms not available on existing visualization websites (E3, E4).
Additionally, the ability to animate incorrect code was seen as beneficial for experimental learning. 
E3 emphasized, \textit{``Being able to run incorrect code and immediately see the consequences through visualization encourages students to experiment and understand mistakes.''}

\textbf{\textit{Enhancing Student Engagement}: }
Several experts noted that student disengagement remains a persistent challenge in DSA education and viewed our system as a promising tool for fostering interest among students who struggle with motivation. E1 observed that \textit{``interactive visualization can promote active learning and increase students' interest,''} while E3 added, \textit{``when I play the animation, they always pay attention to it.''} This feedback highlights the system's potential to shift learning from passive consumption of static materials to active engagement through interactive animation.

\textbf{\textit{Need for Supplemental Explanations}: }
Finally, several experts identified limitations related to explainability. Although the system relies on pseudocode to represent algorithmic processes at each step, pseudocode can be overly abstract for some learners (e.g., a classical algorithm such as Dijkstra's shortest path has concise pseudocode but a complex implementation). As a result, students may struggle to interpret what the animations represent and why specific actions occur when relying solely on pseudocode.
To address this issue, some experts recommended enhancements such as detailed explanatory text or audio annotations for each action (E4, E6), as well as features that highlight corresponding textbook sections explaining the current algorithmic steps (E3). These additions could help bridge the gap between visual representations and the underlying algorithmic concepts.
\section{Discussion}
\label{sec:discussion}

\sys represents a novel tool that addresses salient challenges in AV by automatically generating LLM-powered visualizations from static pseudocode directly on curriculum learning materials. Based on our evaluations, the system achieves the design goals described in \Cref{sec:design-challenges-goals}:
\circledVarD{G1} by introducing an AV Code Generator that produces robust visualization code using multiple LLM agents with reflection mechanisms;
\circledVarD{G2} by displaying the AV panel directly on textbook content, with synchronized code-line highlighting and variable-tracing overlays;
\circledVarD{G3} by consolidating algorithm visualization, time/space complexity visualization, and code editing in a unified panel to ``provide multiple views,'' while offering intuitive animation controllers for ``flexible execution control'' and overlays that ``complement visualizations with explanations'';
\circledVarD{G4} by supporting both Python and JavaScript within a scalable framework; and
\circledVarD{G5} by enabling dynamic input modification and code editing.
Below, we further discuss and reflect on some of the lessons learned and outline opportunities for future research in this area, based on this work.


\textbf{\textit{Towards Classroom Deployments}:}
As represented by the design goals, the primary focus of this work was assessing and validating the UI/UX aspects of \sys, primarily via a controlled usability study and supplemental expert feedback from interviews. As mentioned in \Cref{sec:usability_study}, this usability focus was the reason that recruited student participants needed prior DSA exposure, which aligns with other examples of studies that first focus on the design of learning-focused tools (e.g.,~\cite{vizcoach, gunturu_augmented_2024a, chulpongsatorn_augmented_2023}). Given the success in these validations, in the future we intend to deploy \sys in classroom settings and conduct longitudinal studies of how the tool supports students, such as its impacts on learning outcomes, engagement, and knowledge retention across varying levels of prior experience.

\textbf{\textit{Broadening Support for DSA Topics and Techniques}:}
In this paper, we demonstrated that \sys effectively supports core algorithms and data structures commonly taught in university curricula. However, many DSA courses also cover more advanced topics, such as graph algorithms and red-black trees. Supporting these would require extending the visualization components and prompts for new algorithm types. 
Moreover, the current system supports only Python and JavaScript, limiting its use in courses that rely on other languages (e.g., C++, Java). As the system employs a modular design, both algorithm coverage and language support can easily be extended without major architectural changes.
Additionally, though the current system primarily targets introductory learners, adapting explanations and visual representations to different expertise levels (e.g., beginner, intermediate, advanced) could also broaden its applicability. Such adaptability could enable more personalized learning experiences and enhance the system’s pedagogical impact across diverse educational contexts.

\textbf{\textit{Verifying LLM-Generated AV}:}
Our technical evaluation confirmed that the system generates AV with high robustness. However, as with any LLM-driven tool, there is always the potential for off-target and/or hallucinated results; for instance, the \textit{Aesthetics} sometimes fail to meet expectations. Such risks are widely acknowledged across the education and learning sciences communities~\cite{van2024chatgpt}.  Future work here will add mitigation features to reduce such possibilities: e.g., a reflection mechanism could assess visualization validity and consistency and trigger regeneration when issues are detected, and confidence scores could alert students to consult an instructor when confidence is low. Such features could help balance the risks of misinterpretation while maintaining usability for self-directed learning.

\textbf{\textit{More Intuitive Visual Representations}:}
Although the generated animations effectively demonstrate algorithmic processes, several experts suggested ways to make them more explanatory. For example, E5 remarked, \textit{`Adding visual effects for each loop, or clearly distinguishing between nested and outer loops, would make it easier for students to grasp the differences between algorithms.''} Additionally, E6 proposed incorporating well-known metaphors (e.g., bubble animations for bubble sort) to strengthen the connection between algorithm names, behaviors, and their visual representations. Such additions could more clearly emphasize structural elements and the current DSA concepts of interest, though this would increase the complexity and learning curve of the overall UI/UX, and would thus need to be carefully designed.
One strategy here could be to support multiple levels of abstraction and explanation (e.g., a ``minimal'' visualization style vs. an embellished or heavily annotated style), similar to approaches supported by tools such as \textit{Chisel}~\cite{hayatpur_shapes_2025}. Recent LLM-based approaches for automatically generating stylized infographics~\cite{dibia_lida_2023, xiao_let_2024, zhou_epigraphics_2024} also suggest opportunities to create more expressive and personalizable visualizations.

\textbf{\textit{More In-Depth Understanding}:}
Although the user study revealed that our system is appropriately designed for facilitating learners in understanding DSA concepts, some experts (E3, E5, E6) expressed doubts about whether ``simply visualizing algorithms'' ensures true understanding, and they suggested ways to promote deeper learning and engagement. For example, they suggested incorporating interactive assessment features, such as pop quizzes, fill-in-the-blank exercises, and error correction tasks, after the learning phase. In part, we argue that \sys is positioned as a helper to existing curricula and learning materials, which hopefully already include these types of activities. That said, this recommendation also aligns with the best practice of ``supporting dynamic questions''~\cite{naps_exploring_2002} such as periodically presenting short questions to encourage learners to reflect on a visualization. Such features could be integrated into \sys to encourage students to actively engage with a displayed algorithm beyond reviewing and animating it, promoting deeper comprehension through hands-on practice. 
Future work could, for example, explore implementing LLM-powered quiz generation modules to automatically create personalized assessments based on the extracted pseudocode.

\section{Conclusion}
Algorithm visualization tools are pedagogically effective for learning DSA. However, they remain underutilized for several reasons, such as the difficulty in integrating such tools into existing curricula and classroom practices. In this paper, we introduce \sys, an LLM-assisted tool that transforms static pseudocode into interactive algorithm visualizations to support hands-on interaction and learning. By orchestrating LLMs, our system automatically generates visualizations from textbook and other static content. A trio of evaluations, including a usability study with 15 student participants and expert interviews with 6 instructors, indicates that our approach can lower the barriers to creating and accessing DSA-focused visualizations, while supporting more engaging and personalized learning experiences. Looking ahead, extending the system to more complex algorithms and diverse learning environments offers promising directions for future work.

\bibliographystyle{ACM-Reference-Format}
\bibliography{references, others}


\begin{thebibliography}{58}


\ifx \showCODEN    \undefined \def \showCODEN     #1{\unskip}     \fi
\ifx \showISBNx    \undefined \def \showISBNx     #1{\unskip}     \fi
\ifx \showISBNxiii \undefined \def \showISBNxiii  #1{\unskip}     \fi
\ifx \showISSN     \undefined \def \showISSN      #1{\unskip}     \fi
\ifx \showLCCN     \undefined \def \showLCCN      #1{\unskip}     \fi
\ifx \shownote     \undefined \def \shownote      #1{#1}          \fi
\ifx \showarticletitle \undefined \def \showarticletitle #1{#1}   \fi
\ifx \showURL      \undefined \def \showURL       {\relax}        \fi
\providecommand\bibfield[2]{#2}
\providecommand\bibinfo[2]{#2}
\providecommand\natexlab[1]{#1}
\providecommand\showeprint[2][]{arXiv:#2}

\bibitem[noa(2018)]%
        {noauthor_algorithm_2018}
 \bibinfo{year}{2018}\natexlab{}.
\newblock \bibinfo{title}{Algorithm {Visualizer}}.
\newblock
\urldef\tempurl%
\url{https://algorithm-visualizer.org/}
\showURL{%
\tempurl}


\bibitem[Badam et~al\mbox{.}(2019)]%
        {badam_elastic_2019}
\bibfield{author}{\bibinfo{person}{Sriram~Karthik Badam}, \bibinfo{person}{Zhicheng Liu}, {and} \bibinfo{person}{Niklas Elmqvist}.} \bibinfo{year}{2019}\natexlab{}.
\newblock \showarticletitle{Elastic {Documents}: {Coupling} {Text} and {Tables} through {Contextual} {Visualizations} for {Enhanced} {Document} {Reading}}.
\newblock \bibinfo{journal}{\emph{IEEE Transactions on Visualization and Computer Graphics}} \bibinfo{volume}{25}, \bibinfo{number}{1} (\bibinfo{date}{Jan.} \bibinfo{year}{2019}), \bibinfo{pages}{661--671}.
\newblock
\showISSN{1941-0506}
\href{https://doi.org/10.1109/TVCG.2018.2865119}{doi:\nolinkurl{10.1109/TVCG.2018.2865119}}


\bibitem[Ben-Ari et~al\mbox{.}(2001)]%
        {ben-ari_perspectives_2001}
\bibfield{author}{\bibinfo{person}{Mordechai Ben-Ari}, \bibinfo{person}{Niko Myller}, \bibinfo{person}{Erkki Sutinen}, {and} \bibinfo{person}{Jorma Tarhio}.} \bibinfo{year}{2001}\natexlab{}.
\newblock \showarticletitle{Perspectives on {Program} {Animation} with {Jeliot}}. In \bibinfo{booktitle}{\emph{Revised {Lectures} on {Software} {Visualization}, {International} {Seminar}}}. \bibinfo{publisher}{Springer-Verlag}, \bibinfo{address}{Berlin, Heidelberg}, \bibinfo{pages}{31--45}.
\newblock
\showISBNx{978-3-540-43323-1}


\bibitem[Bostock et~al\mbox{.}(2011)]%
        {bostock_d3_2011}
\bibfield{author}{\bibinfo{person}{Michael Bostock}, \bibinfo{person}{Vadim Ogievetsky}, {and} \bibinfo{person}{Jeffrey Heer}.} \bibinfo{year}{2011}\natexlab{}.
\newblock \showarticletitle{D³ {Data}-{Driven} {Documents}}.
\newblock \bibinfo{journal}{\emph{IEEE Transactions on Visualization and Computer Graphics}} \bibinfo{volume}{17}, \bibinfo{number}{12} (\bibinfo{date}{Dec.} \bibinfo{year}{2011}), \bibinfo{pages}{2301--2309}.
\newblock
\showISSN{1941-0506}
\href{https://doi.org/10.1109/TVCG.2011.185}{doi:\nolinkurl{10.1109/TVCG.2011.185}}


\bibitem[Brooke(1995)]%
        {john_sus_1995}
\bibfield{author}{\bibinfo{person}{John Brooke}.} \bibinfo{year}{1995}\natexlab{}.
\newblock \bibinfo{title}{{SUS} -- a quick and dirty usability scale}.
\newblock
\urldef\tempurl%
\url{https://www.researchgate.net/publication/319394819_SUS_--_a_quick_and_dirty_usability_scale}
\showURL{%
\tempurl}


\bibitem[Brown(1988)]%
        {brown_algorithm_1988}
\bibfield{author}{\bibinfo{person}{Marc~H. Brown}.} \bibinfo{year}{1988}\natexlab{}.
\newblock \bibinfo{booktitle}{\emph{Algorithm {Animation}}}.
\newblock \bibinfo{publisher}{MIT Press}, \bibinfo{address}{Cambridge, MA, USA}.
\newblock
\showISBNx{978-0-262-52411-7}


\bibitem[Brown and Sedgewick(1984)]%
        {brown_system_1984}
\bibfield{author}{\bibinfo{person}{Marc~H. Brown} {and} \bibinfo{person}{Robert Sedgewick}.} \bibinfo{year}{1984}\natexlab{}.
\newblock \showarticletitle{A system for algorithm animation}.
\newblock \bibinfo{journal}{\emph{SIGGRAPH Comput. Graph.}} \bibinfo{volume}{18}, \bibinfo{number}{3} (\bibinfo{date}{Jan.} \bibinfo{year}{1984}), \bibinfo{pages}{177--186}.
\newblock
\showISSN{0097-8930}
\href{https://doi.org/10.1145/964965.808596}{doi:\nolinkurl{10.1145/964965.808596}}


\bibitem[Chawla et~al\mbox{.}(2026)]%
        {vizcoach}
\bibfield{author}{\bibinfo{person}{S. Chawla}, \bibinfo{person}{M. Kintscher}, \bibinfo{person}{J. Narula}, \bibinfo{person}{A. Arunkumar}, \bibinfo{person}{A. Amresh}, {and} \bibinfo{person}{C. Bryan}.} \bibinfo{year}{2026}\natexlab{}.
\newblock \showarticletitle{VizCoach: Designing an Orchestration-Based Tool for Data Visualization Education}. In \bibinfo{booktitle}{\emph{Proceedings of the 59th Hawaii International Conference on System Sciences (HICSS)}}.
\newblock


\bibitem[Chen et~al\mbox{.}(2023)]%
        {chen_marvista_2023}
\bibfield{author}{\bibinfo{person}{Xiang~“Anthony” Chen}, \bibinfo{person}{Chien-Sheng Wu}, \bibinfo{person}{Lidiya Murakhovs’ka}, \bibinfo{person}{Philippe Laban}, \bibinfo{person}{Tong Niu}, \bibinfo{person}{Wenhao Liu}, {and} \bibinfo{person}{Caiming Xiong}.} \bibinfo{year}{2023}\natexlab{}.
\newblock \showarticletitle{Marvista: {Exploring} the {Design} of a {Human}-{AI} {Collaborative} {News} {Reading} {Tool}}.
\newblock \bibinfo{journal}{\emph{ACM Trans. Comput.-Hum. Interact.}} \bibinfo{volume}{30}, \bibinfo{number}{6} (\bibinfo{date}{Sept.} \bibinfo{year}{2023}), \bibinfo{pages}{92:1--92:27}.
\newblock
\showISSN{1073-0516}
\href{https://doi.org/10.1145/3609331}{doi:\nolinkurl{10.1145/3609331}}


\bibitem[Chi et~al\mbox{.}(2005)]%
        {chi_scenthighlights_2005}
\bibfield{author}{\bibinfo{person}{Ed~H. Chi}, \bibinfo{person}{Lichan Hong}, \bibinfo{person}{Michelle Gumbrecht}, {and} \bibinfo{person}{Stuart~K. Card}.} \bibinfo{year}{2005}\natexlab{}.
\newblock \showarticletitle{{ScentHighlights}: highlighting conceptually-related sentences during reading}. In \bibinfo{booktitle}{\emph{Proceedings of the 10th international conference on {Intelligent} user interfaces}} \emph{(\bibinfo{series}{{IUI} '05})}. \bibinfo{publisher}{Association for Computing Machinery}, \bibinfo{address}{New York, NY, USA}, \bibinfo{pages}{272--274}.
\newblock
\showISBNx{978-1-58113-894-8}
\href{https://doi.org/10.1145/1040830.1040895}{doi:\nolinkurl{10.1145/1040830.1040895}}


\bibitem[Chulpongsatorn et~al\mbox{.}(2023)]%
        {chulpongsatorn_augmented_2023}
\bibfield{author}{\bibinfo{person}{Neil Chulpongsatorn}, \bibinfo{person}{Mille~Skovhus Lunding}, \bibinfo{person}{Nishan Soni}, {and} \bibinfo{person}{Ryo Suzuki}.} \bibinfo{year}{2023}\natexlab{}.
\newblock \showarticletitle{Augmented {Math}: {Authoring} {AR}-{Based} {Explorable} {Explanations} by {Augmenting} {Static} {Math} {Textbooks}}. In \bibinfo{booktitle}{\emph{Proceedings of the 36th {Annual} {ACM} {Symposium} on {User} {Interface} {Software} and {Technology}}} \emph{(\bibinfo{series}{{UIST} '23})}. \bibinfo{publisher}{Association for Computing Machinery}, \bibinfo{address}{New York, NY, USA}, \bibinfo{pages}{1--16}.
\newblock
\showISBNx{979-8-4007-0132-0}
\href{https://doi.org/10.1145/3586183.3606827}{doi:\nolinkurl{10.1145/3586183.3606827}}


\bibitem[Concepcion et~al\mbox{.}(2000)]%
        {concepcion_algorithma_2000}
\bibfield{author}{\bibinfo{person}{Arturo~I. Concepcion}, \bibinfo{person}{Nathan Leach}, {and} \bibinfo{person}{Allan Knight}.} \bibinfo{year}{2000}\natexlab{}.
\newblock \showarticletitle{Algorithma 99: an experiment in reusability \& component based software engineering}.
\newblock \bibinfo{journal}{\emph{SIGCSE Bull.}} \bibinfo{volume}{32}, \bibinfo{number}{1} (\bibinfo{date}{March} \bibinfo{year}{2000}), \bibinfo{pages}{162--166}.
\newblock
\showISSN{0097-8418}
\href{https://doi.org/10.1145/331795.331847}{doi:\nolinkurl{10.1145/331795.331847}}


\bibitem[Cormen et~al\mbox{.}(2009)]%
        {cormen2009introduction}
\bibfield{author}{\bibinfo{person}{Thomas~H. Cormen}, \bibinfo{person}{Charles~E. Leiserson}, \bibinfo{person}{Ronald~L. Rivest}, {and} \bibinfo{person}{Clifford Stein}.} \bibinfo{year}{2009}\natexlab{}.
\newblock \bibinfo{booktitle}{\emph{Introduction to Algorithms, Third Edition} (\bibinfo{edition}{3rd} ed.)}.
\newblock \bibinfo{publisher}{The MIT Press}.
\newblock
\showISBNx{0262033844}


\bibitem[{David Galles}(2011a)]%
        {david_galles_cs_2011}
\bibfield{author}{\bibinfo{person}{{David Galles}}.} \bibinfo{year}{2011}\natexlab{a}.
\newblock \bibinfo{title}{{CS} 1332 {Data} {Structures} and {Algorithms} {Visualizations}}.
\newblock
\urldef\tempurl%
\url{https://csvistool.com/}
\showURL{%
\tempurl}


\bibitem[{David Galles}(2011b)]%
        {david_galles_data_2011}
\bibfield{author}{\bibinfo{person}{{David Galles}}.} \bibinfo{year}{2011}\natexlab{b}.
\newblock \bibinfo{title}{Data {Structure} {Visualization}}.
\newblock
\urldef\tempurl%
\url{https://www.cs.usfca.edu/~galles/visualization/about.html}
\showURL{%
\tempurl}


\bibitem[Dibia(2023)]%
        {dibia_lida_2023}
\bibfield{author}{\bibinfo{person}{Victor Dibia}.} \bibinfo{year}{2023}\natexlab{}.
\newblock \showarticletitle{{LIDA}: {A} {Tool} for {Automatic} {Generation} of {Grammar}-{Agnostic} {Visualizations} and {Infographics} using {Large} {Language} {Models}}. In \bibinfo{booktitle}{\emph{Proceedings of the 61st {Annual} {Meeting} of the {Association} for {Computational} {Linguistics} ({Volume} 3: {System} {Demonstrations})}}, \bibfield{editor}{\bibinfo{person}{Danushka Bollegala}, \bibinfo{person}{Ruihong Huang}, {and} \bibinfo{person}{Alan Ritter}} (Eds.). \bibinfo{publisher}{Association for Computational Linguistics}, \bibinfo{address}{Toronto, Canada}, \bibinfo{pages}{113--126}.
\newblock
\href{https://doi.org/10.18653/v1/2023.acl-demo.11}{doi:\nolinkurl{10.18653/v1/2023.acl-demo.11}}


\bibitem[Fok et~al\mbox{.}(2023)]%
        {fok_scim_2023}
\bibfield{author}{\bibinfo{person}{Raymond Fok}, \bibinfo{person}{Hita Kambhamettu}, \bibinfo{person}{Luca Soldaini}, \bibinfo{person}{Jonathan Bragg}, \bibinfo{person}{Kyle Lo}, \bibinfo{person}{Marti Hearst}, \bibinfo{person}{Andrew Head}, {and} \bibinfo{person}{Daniel~S Weld}.} \bibinfo{year}{2023}\natexlab{}.
\newblock \showarticletitle{Scim: {Intelligent} {Skimming} {Support} for {Scientific} {Papers}}. In \bibinfo{booktitle}{\emph{Proceedings of the 28th {International} {Conference} on {Intelligent} {User} {Interfaces}}} \emph{(\bibinfo{series}{{IUI} '23})}. \bibinfo{publisher}{Association for Computing Machinery}, \bibinfo{address}{New York, NY, USA}, \bibinfo{pages}{476--490}.
\newblock
\showISBNx{979-8-4007-0106-1}
\href{https://doi.org/10.1145/3581641.3584034}{doi:\nolinkurl{10.1145/3581641.3584034}}


\bibitem[Fouh et~al\mbox{.}(2012)]%
        {fouh_role_2012}
\bibfield{author}{\bibinfo{person}{Eric Fouh}, \bibinfo{person}{Monika Akbar}, {and} \bibinfo{person}{Clifford~A. Shaffer}.} \bibinfo{year}{2012}\natexlab{}.
\newblock \showarticletitle{The {Role} of {Visualization} in {Computer} {Science} {Education}}.
\newblock \bibinfo{journal}{\emph{Computers in the Schools}} \bibinfo{volume}{29}, \bibinfo{number}{1-2} (\bibinfo{date}{Jan.} \bibinfo{year}{2012}), \bibinfo{pages}{95--117}.
\newblock
\showISSN{0738-0569, 1528-7033}
\href{https://doi.org/10.1080/07380569.2012.651422}{doi:\nolinkurl{10.1080/07380569.2012.651422}}


\bibitem[Goel et~al\mbox{.}(2023)]%
        {goel_review_2023}
\bibfield{author}{\bibinfo{person}{Sarthak Goel}, \bibinfo{person}{Vanshika Varshney}, \bibinfo{person}{Shubham Dikshant}, \bibinfo{person}{Akhil Sharma}, {and} \bibinfo{person}{Sherish Johri}.} \bibinfo{year}{2023}\natexlab{}.
\newblock \showarticletitle{A {Review} of {The} {Algorithm} {Visualization} {Field}}. In \bibinfo{booktitle}{\emph{2023 14th {International} {Conference} on {Computing} {Communication} and {Networking} {Technologies} ({ICCCNT})}}. \bibinfo{pages}{1--5}.
\newblock
\showISSN{2473-7674}
\href{https://doi.org/10.1109/ICCCNT56998.2023.10307685}{doi:\nolinkurl{10.1109/ICCCNT56998.2023.10307685}}


\bibitem[Grissom et~al\mbox{.}(2003)]%
        {grissom_algorithm_2003}
\bibfield{author}{\bibinfo{person}{Scott Grissom}, \bibinfo{person}{Myles~F. McNally}, {and} \bibinfo{person}{Tom Naps}.} \bibinfo{year}{2003}\natexlab{}.
\newblock \showarticletitle{Algorithm visualization in {CS} education: comparing levels of student engagement}. In \bibinfo{booktitle}{\emph{Proceedings of the 2003 {ACM} symposium on {Software} visualization}} \emph{(\bibinfo{series}{{SoftVis} '03})}. \bibinfo{publisher}{Association for Computing Machinery}, \bibinfo{address}{New York, NY, USA}, \bibinfo{pages}{87--94}.
\newblock
\showISBNx{978-1-58113-642-5}
\href{https://doi.org/10.1145/774833.774846}{doi:\nolinkurl{10.1145/774833.774846}}


\bibitem[Gunturu et~al\mbox{.}(2024a)]%
        {gunturu_augmented_2024}
\bibfield{author}{\bibinfo{person}{Aditya Gunturu}, \bibinfo{person}{Yi Wen}, \bibinfo{person}{Nandi Zhang}, \bibinfo{person}{Jarin Thundathil}, \bibinfo{person}{Rubaiat~Habib Kazi}, {and} \bibinfo{person}{Ryo Suzuki}.} \bibinfo{year}{2024}\natexlab{a}.
\newblock \showarticletitle{Augmented {Physics}: {Creating} {Interactive} and {Embedded} {Physics} {Simulations} from {Static} {Textbook} {Diagrams}}. In \bibinfo{booktitle}{\emph{Proceedings of the 37th {Annual} {ACM} {Symposium} on {User} {Interface} {Software} and {Technology}}} \emph{(\bibinfo{series}{{UIST} '24})}. \bibinfo{publisher}{Association for Computing Machinery}, \bibinfo{address}{New York, NY, USA}, \bibinfo{pages}{1--12}.
\newblock
\showISBNx{979-8-4007-0628-8}
\href{https://doi.org/10.1145/3654777.3676392}{doi:\nolinkurl{10.1145/3654777.3676392}}


\bibitem[Gunturu et~al\mbox{.}(2024b)]%
        {gunturu_augmented_2024a}
\bibfield{author}{\bibinfo{person}{Aditya Gunturu}, \bibinfo{person}{Yi Wen}, \bibinfo{person}{Nandi Zhang}, \bibinfo{person}{Jarin Thundathil}, \bibinfo{person}{Rubaiat~Habib Kazi}, {and} \bibinfo{person}{Ryo Suzuki}.} \bibinfo{year}{2024}\natexlab{b}.
\newblock \showarticletitle{Augmented {Physics}: {Creating} {Interactive} and {Embedded} {Physics} {Simulations} from {Static} {Textbook} {Diagrams}}. In \bibinfo{booktitle}{\emph{Proceedings of the 37th {Annual} {ACM} {Symposium} on {User} {Interface} {Software} and {Technology}}} \emph{(\bibinfo{series}{{UIST} '24})}. \bibinfo{publisher}{Association for Computing Machinery}, \bibinfo{address}{New York, NY, USA}, \bibinfo{pages}{1--12}.
\newblock
\showISBNx{979-8-4007-0628-8}
\href{https://doi.org/10.1145/3654777.3676392}{doi:\nolinkurl{10.1145/3654777.3676392}}


\bibitem[Guo(2013)]%
        {guo_online_2013}
\bibfield{author}{\bibinfo{person}{Philip~J. Guo}.} \bibinfo{year}{2013}\natexlab{}.
\newblock \showarticletitle{Online python tutor: embeddable web-based program visualization for cs education}. In \bibinfo{booktitle}{\emph{Proceeding of the 44th {ACM} technical symposium on {Computer} science education}}. \bibinfo{publisher}{ACM}, \bibinfo{address}{Denver Colorado USA}, \bibinfo{pages}{579--584}.
\newblock
\showISBNx{978-1-4503-1868-6}
\href{https://doi.org/10.1145/2445196.2445368}{doi:\nolinkurl{10.1145/2445196.2445368}}


\bibitem[Hall et~al\mbox{.}(2013)]%
        {hall_evaluating_2013}
\bibfield{author}{\bibinfo{person}{Simin Hall}, \bibinfo{person}{Clifford Shaffer}, \bibinfo{person}{Eric Fouh}, \bibinfo{person}{Mai ElShehaly}, {and} \bibinfo{person}{Daniel Breakiron}.} \bibinfo{year}{2013}\natexlab{}.
\newblock \showarticletitle{Evaluating {Online} {Tutorials} for {Data} {Structures} and {Algorithms} {Courses}}. In \bibinfo{booktitle}{\emph{2013 {ASEE} {Annual} {Conference} \& {Exposition} {Proceedings}}}. \bibinfo{publisher}{ASEE Conferences}, \bibinfo{address}{Atlanta, Georgia}, \bibinfo{pages}{23.549.1--23.549.14}.
\newblock
\href{https://doi.org/10.18260/1-2--19563}{doi:\nolinkurl{10.18260/1-2--19563}}


\bibitem[Hansen et~al\mbox{.}(2002)]%
        {hansen_designing_2002}
\bibfield{author}{\bibinfo{person}{Steven Hansen}, \bibinfo{person}{N.~Hari Narayanan}, {and} \bibinfo{person}{Mary Hegarty}.} \bibinfo{year}{2002}\natexlab{}.
\newblock \showarticletitle{Designing {Educationally} {Effective} {Algorithm} {Visualizations}}.
\newblock \bibinfo{journal}{\emph{Journal of Visual Languages \& Computing}} \bibinfo{volume}{13}, \bibinfo{number}{3} (\bibinfo{date}{June} \bibinfo{year}{2002}), \bibinfo{pages}{291--317}.
\newblock
\showISSN{1045-926X}
\href{https://doi.org/10.1006/jvlc.2002.0236}{doi:\nolinkurl{10.1006/jvlc.2002.0236}}


\bibitem[Hayatpur et~al\mbox{.}(2025)]%
        {hayatpur_shapes_2025}
\bibfield{author}{\bibinfo{person}{Devamardeep Hayatpur}, \bibinfo{person}{Brian Hempel}, \bibinfo{person}{Richard Lin}, {and} \bibinfo{person}{Haijun Xia}.} \bibinfo{year}{2025}\natexlab{}.
\newblock \showarticletitle{The {Shapes} of {Abstraction} in {Data} {Structure} {Diagrams}}. In \bibinfo{booktitle}{\emph{Proceedings of the 2025 {CHI} {Conference} on {Human} {Factors} in {Computing} {Systems}}} \emph{(\bibinfo{series}{{CHI} '25})}. \bibinfo{publisher}{Association for Computing Machinery}, \bibinfo{address}{New York, NY, USA}, \bibinfo{pages}{1--12}.
\newblock
\showISBNx{979-8-4007-1394-1}
\href{https://doi.org/10.1145/3706598.3713723}{doi:\nolinkurl{10.1145/3706598.3713723}}


\bibitem[Hullman et~al\mbox{.}(2013)]%
        {hullman_contextifier_2013}
\bibfield{author}{\bibinfo{person}{Jessica Hullman}, \bibinfo{person}{Nicholas Diakopoulos}, {and} \bibinfo{person}{Eytan Adar}.} \bibinfo{year}{2013}\natexlab{}.
\newblock \showarticletitle{Contextifier: automatic generation of annotated stock visualizations}. In \bibinfo{booktitle}{\emph{Proceedings of the {SIGCHI} {Conference} on {Human} {Factors} in {Computing} {Systems}}} \emph{(\bibinfo{series}{{CHI} '13})}. \bibinfo{publisher}{Association for Computing Machinery}, \bibinfo{address}{New York, NY, USA}, \bibinfo{pages}{2707--2716}.
\newblock
\showISBNx{978-1-4503-1899-0}
\href{https://doi.org/10.1145/2470654.2481374}{doi:\nolinkurl{10.1145/2470654.2481374}}


\bibitem[Hundhausen and Brown(2008)]%
        {hudson_designing_2008}
\bibfield{author}{\bibinfo{person}{Christopher~D. Hundhausen} {and} \bibinfo{person}{Jonathan~L. Brown}.} \bibinfo{year}{2008}\natexlab{}.
\newblock \showarticletitle{Designing, visualizing, and discussing algorithms within a CS 1 studio experience: An empirical study}.
\newblock \bibinfo{journal}{\emph{Computers \& Education}} \bibinfo{volume}{50}, \bibinfo{number}{1} (\bibinfo{year}{2008}), \bibinfo{pages}{301--326}.
\newblock
\showISSN{0360-1315}
\href{https://doi.org/10.1016/j.compedu.2006.06.002}{doi:\nolinkurl{10.1016/j.compedu.2006.06.002}}


\bibitem[Hundhausen et~al\mbox{.}(2002)]%
        {hundhausen_metastudy_2002}
\bibfield{author}{\bibinfo{person}{Christopher~D. Hundhausen}, \bibinfo{person}{Sarah~A. Douglas}, {and} \bibinfo{person}{John~T. Stasko}.} \bibinfo{year}{2002}\natexlab{}.
\newblock \showarticletitle{A {Meta}-{Study} of {Algorithm} {Visualization} {Effectiveness}}.
\newblock \bibinfo{journal}{\emph{Journal of Visual Languages \& Computing}} \bibinfo{volume}{13}, \bibinfo{number}{3} (\bibinfo{date}{June} \bibinfo{year}{2002}), \bibinfo{pages}{259--290}.
\newblock
\showISSN{1045-926X}
\href{https://doi.org/10.1006/jvlc.2002.0237}{doi:\nolinkurl{10.1006/jvlc.2002.0237}}


\bibitem[Isohanni and Järvinen(2014)]%
        {isohanni_are_2014}
\bibfield{author}{\bibinfo{person}{Essi Isohanni} {and} \bibinfo{person}{Hannu-Matti Järvinen}.} \bibinfo{year}{2014}\natexlab{}.
\newblock \showarticletitle{Are visualization tools used in programming education? by whom, how, why, and why not?}. In \bibinfo{booktitle}{\emph{Proceedings of the 14th {Koli} {Calling} {International} {Conference} on {Computing} {Education} {Research}}} \emph{(\bibinfo{series}{Koli {Calling} '14})}. \bibinfo{publisher}{Association for Computing Machinery}, \bibinfo{address}{New York, NY, USA}, \bibinfo{pages}{35--40}.
\newblock
\showISBNx{978-1-4503-3065-7}
\href{https://doi.org/10.1145/2674683.2674688}{doi:\nolinkurl{10.1145/2674683.2674688}}


\bibitem[Kang et~al\mbox{.}(2022)]%
        {kang_threddy_2022}
\bibfield{author}{\bibinfo{person}{Hyeonsu Kang}, \bibinfo{person}{Joseph~Chee Chang}, \bibinfo{person}{Yongsung Kim}, {and} \bibinfo{person}{Aniket Kittur}.} \bibinfo{year}{2022}\natexlab{}.
\newblock \showarticletitle{Threddy: {An} {Interactive} {System} for {Personalized} {Thread}-based {Exploration} and {Organization} of {Scientific} {Literature}}. In \bibinfo{booktitle}{\emph{Proceedings of the 35th {Annual} {ACM} {Symposium} on {User} {Interface} {Software} and {Technology}}} \emph{(\bibinfo{series}{{UIST} '22})}. \bibinfo{publisher}{Association for Computing Machinery}, \bibinfo{address}{New York, NY, USA}, \bibinfo{pages}{1--15}.
\newblock
\showISBNx{978-1-4503-9320-1}
\href{https://doi.org/10.1145/3526113.3545660}{doi:\nolinkurl{10.1145/3526113.3545660}}


\bibitem[Kehoe et~al\mbox{.}(2001)]%
        {kehoe_rethinking_2001}
\bibfield{author}{\bibinfo{person}{Colleen Kehoe}, \bibinfo{person}{John Stasko}, {and} \bibinfo{person}{Ashley Taylor}.} \bibinfo{year}{2001}\natexlab{}.
\newblock \showarticletitle{Rethinking the evaluation of algorithm animations as learning aids: an observational study}.
\newblock \bibinfo{journal}{\emph{International Journal of Human-Computer Studies}} \bibinfo{volume}{54}, \bibinfo{number}{2} (\bibinfo{date}{Feb.} \bibinfo{year}{2001}), \bibinfo{pages}{265--284}.
\newblock
\showISSN{1071-5819}
\href{https://doi.org/10.1006/ijhc.2000.0409}{doi:\nolinkurl{10.1006/ijhc.2000.0409}}


\bibitem[Kerren and Stasko(2001)]%
        {kerren_algorithm_2001}
\bibfield{author}{\bibinfo{person}{Andreas Kerren} {and} \bibinfo{person}{John~T. Stasko}.} \bibinfo{year}{2001}\natexlab{}.
\newblock \showarticletitle{Algorithm {Animation} - {Introduction}}. In \bibinfo{booktitle}{\emph{Revised {Lectures} on {Software} {Visualization}, {International} {Seminar}}}. \bibinfo{publisher}{Springer-Verlag}, \bibinfo{address}{Berlin, Heidelberg}, \bibinfo{pages}{1--15}.
\newblock
\showISBNx{978-3-540-43323-1}


\bibitem[Kim et~al\mbox{.}(2018)]%
        {kim_facilitating_2018}
\bibfield{author}{\bibinfo{person}{Dae~Hyun Kim}, \bibinfo{person}{Enamul Hoque}, \bibinfo{person}{Juho Kim}, {and} \bibinfo{person}{Maneesh Agrawala}.} \bibinfo{year}{2018}\natexlab{}.
\newblock \showarticletitle{Facilitating {Document} {Reading} by {Linking} {Text} and {Tables}}. In \bibinfo{booktitle}{\emph{Proceedings of the 31st {Annual} {ACM} {Symposium} on {User} {Interface} {Software} and {Technology}}} \emph{(\bibinfo{series}{{UIST} '18})}. \bibinfo{publisher}{Association for Computing Machinery}, \bibinfo{address}{New York, NY, USA}, \bibinfo{pages}{423--434}.
\newblock
\showISBNx{978-1-4503-5948-1}
\href{https://doi.org/10.1145/3242587.3242617}{doi:\nolinkurl{10.1145/3242587.3242617}}


\bibitem[Li et~al\mbox{.}(2019)]%
        {li_holodoc_2019}
\bibfield{author}{\bibinfo{person}{Zhen Li}, \bibinfo{person}{Michelle Annett}, \bibinfo{person}{Ken Hinckley}, \bibinfo{person}{Karan Singh}, {and} \bibinfo{person}{Daniel Wigdor}.} \bibinfo{year}{2019}\natexlab{}.
\newblock \showarticletitle{{HoloDoc}: {Enabling} {Mixed} {Reality} {Workspaces} that {Harness} {Physical} and {Digital} {Content}}. In \bibinfo{booktitle}{\emph{Proceedings of the 2019 {CHI} {Conference} on {Human} {Factors} in {Computing} {Systems}}} \emph{(\bibinfo{series}{{CHI} '19})}. \bibinfo{publisher}{Association for Computing Machinery}, \bibinfo{address}{New York, NY, USA}, \bibinfo{pages}{1--14}.
\newblock
\showISBNx{978-1-4503-5970-2}
\href{https://doi.org/10.1145/3290605.3300917}{doi:\nolinkurl{10.1145/3290605.3300917}}


\bibitem[Lin and Dwyer(2010)]%
        {lin2010effect}
\bibfield{author}{\bibinfo{person}{Huifen Lin} {and} \bibinfo{person}{Francis Dwyer}.} \bibinfo{year}{2010}\natexlab{}.
\newblock \showarticletitle{The effect of static and animated visualization: a perspective of instructional effectiveness and efficiency}.
\newblock \bibinfo{journal}{\emph{Educational Technology Research and Development}} \bibinfo{volume}{58}, \bibinfo{number}{2} (\bibinfo{year}{2010}), \bibinfo{pages}{155--174}.
\newblock


\bibitem[Masson et~al\mbox{.}(2023)]%
        {masson_charagraph_2023}
\bibfield{author}{\bibinfo{person}{Damien Masson}, \bibinfo{person}{Sylvain Malacria}, \bibinfo{person}{Géry Casiez}, {and} \bibinfo{person}{Daniel Vogel}.} \bibinfo{year}{2023}\natexlab{}.
\newblock \showarticletitle{Charagraph: {Interactive} {Generation} of {Charts} for {Realtime} {Annotation} of {Data}-{Rich} {Paragraphs}}. In \bibinfo{booktitle}{\emph{Proceedings of the 2023 {CHI} {Conference} on {Human} {Factors} in {Computing} {Systems}}} \emph{(\bibinfo{series}{{CHI} '23})}. \bibinfo{publisher}{Association for Computing Machinery}, \bibinfo{address}{New York, NY, USA}, \bibinfo{pages}{1--18}.
\newblock
\showISBNx{978-1-4503-9421-5}
\href{https://doi.org/10.1145/3544548.3581091}{doi:\nolinkurl{10.1145/3544548.3581091}}


\bibitem[Mayer and Moreno(1998)]%
        {mayer_splitattention_1998}
\bibfield{author}{\bibinfo{person}{Richard~E. Mayer} {and} \bibinfo{person}{Roxana Moreno}.} \bibinfo{year}{1998}\natexlab{}.
\newblock \showarticletitle{A split-attention effect in multimedia learning: {Evidence} for dual processing systems in working memory}.
\newblock \bibinfo{journal}{\emph{Journal of Educational Psychology}} \bibinfo{volume}{90}, \bibinfo{number}{2} (\bibinfo{year}{1998}), \bibinfo{pages}{312--320}.
\newblock
\showISSN{1939-2176}
\href{https://doi.org/10.1037/0022-0663.90.2.312}{doi:\nolinkurl{10.1037/0022-0663.90.2.312}}
\newblock
\shownote{Place: US}.


\bibitem[Mornar et~al\mbox{.}(2014)]%
        {mornar_system_2014}
\bibfield{author}{\bibinfo{person}{Jure Mornar}, \bibinfo{person}{Andrina Granić}, {and} \bibinfo{person}{Saša Mladenović}.} \bibinfo{year}{2014}\natexlab{}.
\newblock \showarticletitle{System for automatic generation of algorithm visualizations based on pseudocode interpretation}. In \bibinfo{booktitle}{\emph{Proceedings of the 2014 conference on {Innovation} \& technology in computer science education}} \emph{(\bibinfo{series}{{ITiCSE} '14})}. \bibinfo{publisher}{Association for Computing Machinery}, \bibinfo{address}{New York, NY, USA}, \bibinfo{pages}{27--32}.
\newblock
\showISBNx{978-1-4503-2833-3}
\href{https://doi.org/10.1145/2591708.2591743}{doi:\nolinkurl{10.1145/2591708.2591743}}


\bibitem[Naps et~al\mbox{.}(2002)]%
        {naps_exploring_2002}
\bibfield{author}{\bibinfo{person}{Thomas~L. Naps}, \bibinfo{person}{Guido Rößling}, \bibinfo{person}{Vicki Almstrum}, \bibinfo{person}{Wanda Dann}, \bibinfo{person}{Rudolf Fleischer}, \bibinfo{person}{Chris Hundhausen}, \bibinfo{person}{Ari Korhonen}, \bibinfo{person}{Lauri Malmi}, \bibinfo{person}{Myles McNally}, \bibinfo{person}{Susan Rodger}, {and} \bibinfo{person}{J.~Ángel Velázquez-Iturbide}.} \bibinfo{year}{2002}\natexlab{}.
\newblock \showarticletitle{Exploring the role of visualization and engagement in computer science education}.
\newblock \bibinfo{journal}{\emph{SIGCSE Bull.}} \bibinfo{volume}{35}, \bibinfo{number}{2} (\bibinfo{date}{June} \bibinfo{year}{2002}), \bibinfo{pages}{131--152}.
\newblock
\showISSN{0097-8418}
\href{https://doi.org/10.1145/782941.782998}{doi:\nolinkurl{10.1145/782941.782998}}


\bibitem[Paredes-Velasco et~al\mbox{.}(2023)]%
        {paredes-velasco_augmented_2023}
\bibfield{author}{\bibinfo{person}{Maximiliano Paredes-Velasco}, \bibinfo{person}{J.~Ángel Velázquez-Iturbide}, {and} \bibinfo{person}{Mónica Gómez-Ríos}.} \bibinfo{year}{2023}\natexlab{}.
\newblock \showarticletitle{Augmented reality with algorithm animation and their effect on students’ emotions}.
\newblock \bibinfo{journal}{\emph{Multimedia Tools and Applications}} \bibinfo{volume}{82}, \bibinfo{number}{8} (\bibinfo{date}{March} \bibinfo{year}{2023}), \bibinfo{pages}{11819--11845}.
\newblock
\showISSN{1573-7721}
\href{https://doi.org/10.1007/s11042-022-13679-1}{doi:\nolinkurl{10.1007/s11042-022-13679-1}}
\newblock
\shownote{Company: Springer Distributor: Springer Institution: Springer Label: Springer Number: 8}.


\bibitem[{School of Computing, National University of Singapore} and Halim(2015)]%
        {school_of_computing_national_university_of_singapore_visualgo_2015}
\bibfield{author}{\bibinfo{person}{{School of Computing, National University of Singapore}} {and} \bibinfo{person}{Steven Halim}.} \bibinfo{year}{2015}\natexlab{}.
\newblock \showarticletitle{{VisuAlgo} – {Visualising} {Data} {Structures} and {Algorithms} {Through} {Animation}}.
\newblock \bibinfo{journal}{\emph{OLYMPIADS IN INFORMATICS}}  \bibinfo{volume}{9} (\bibinfo{date}{July} \bibinfo{year}{2015}), \bibinfo{pages}{243--245}.
\newblock
\showISSN{18227732, 23358955}
\href{https://doi.org/10.15388/ioi.2015.20}{doi:\nolinkurl{10.15388/ioi.2015.20}}


\bibitem[Schweitzer et~al\mbox{.}(2007)]%
        {schweitzer_interactive_2007}
\bibfield{author}{\bibinfo{person}{Dino Schweitzer}, \bibinfo{person}{{View Profile}}, \bibinfo{person}{Wayne Brown}, {and} \bibinfo{person}{{View Profile}}.} \bibinfo{year}{2007}\natexlab{}.
\newblock \showarticletitle{Interactive visualization for the active learning classroom}.
\newblock \bibinfo{journal}{\emph{ACM SIGCSE Bulletin}} \bibinfo{volume}{39}, \bibinfo{number}{1} (\bibinfo{date}{March} \bibinfo{year}{2007}), \bibinfo{pages}{208--212}.
\newblock
\href{https://doi.org/10.1145/1227504.1227384}{doi:\nolinkurl{10.1145/1227504.1227384}}


\bibitem[Sedlmair et~al\mbox{.}(2012)]%
        {sedlmair2012design}
\bibfield{author}{\bibinfo{person}{Michael Sedlmair}, \bibinfo{person}{Miriah Meyer}, {and} \bibinfo{person}{Tamara Munzner}.} \bibinfo{year}{2012}\natexlab{}.
\newblock \showarticletitle{Design study methodology: Reflections from the trenches and the stacks}.
\newblock \bibinfo{journal}{\emph{IEEE Transactions on Visualization and Computer Graphics}} \bibinfo{volume}{18}, \bibinfo{number}{12} (\bibinfo{year}{2012}), \bibinfo{pages}{2431--2440}.
\newblock


\bibitem[Shaffer et~al\mbox{.}(2011)]%
        {shaffer_getting_2011}
\bibfield{author}{\bibinfo{person}{Clifford~A. Shaffer}, \bibinfo{person}{Monika Akbar}, \bibinfo{person}{Alexander Joel~D. Alon}, \bibinfo{person}{Michael Stewart}, {and} \bibinfo{person}{Stephen~H. Edwards}.} \bibinfo{year}{2011}\natexlab{}.
\newblock \showarticletitle{Getting algorithm visualizations into the classroom}. In \bibinfo{booktitle}{\emph{Proceedings of the 42nd {ACM} technical symposium on {Computer} science education}} \emph{(\bibinfo{series}{{SIGCSE} '11})}. \bibinfo{publisher}{Association for Computing Machinery}, \bibinfo{address}{New York, NY, USA}, \bibinfo{pages}{129--134}.
\newblock
\showISBNx{978-1-4503-0500-6}
\href{https://doi.org/10.1145/1953163.1953204}{doi:\nolinkurl{10.1145/1953163.1953204}}


\bibitem[Shaffer et~al\mbox{.}(2010)]%
        {shaffer_algorithm_2010}
\bibfield{author}{\bibinfo{person}{Clifford~A. Shaffer}, \bibinfo{person}{Matthew~L. Cooper}, \bibinfo{person}{Alexander Joel~D. Alon}, \bibinfo{person}{Monika Akbar}, \bibinfo{person}{Michael Stewart}, \bibinfo{person}{Sean Ponce}, {and} \bibinfo{person}{Stephen~H. Edwards}.} \bibinfo{year}{2010}\natexlab{}.
\newblock \showarticletitle{Algorithm {Visualization}: {The} {State} of the {Field}}.
\newblock \bibinfo{journal}{\emph{ACM Trans. Comput. Educ.}} \bibinfo{volume}{10}, \bibinfo{number}{3} (\bibinfo{date}{Aug.} \bibinfo{year}{2010}), \bibinfo{pages}{9:1--9:22}.
\newblock
\href{https://doi.org/10.1145/1821996.1821997}{doi:\nolinkurl{10.1145/1821996.1821997}}


\bibitem[Shen et~al\mbox{.}(2021)]%
        {shen_layoutparser_2021}
\bibfield{author}{\bibinfo{person}{Zejiang Shen}, \bibinfo{person}{Ruochen Zhang}, \bibinfo{person}{Melissa Dell}, \bibinfo{person}{Benjamin Charles~Germain Lee}, \bibinfo{person}{Jacob Carlson}, {and} \bibinfo{person}{Weining Li}.} \bibinfo{year}{2021}\natexlab{}.
\newblock \showarticletitle{{LayoutParser}: {A} {Unified} {Toolkit} for {Deep} {Learning} {Based} {Document} {Image} {Analysis}}. In \bibinfo{booktitle}{\emph{Document {Analysis} and {Recognition} – {ICDAR} 2021: 16th {International} {Conference}, {Lausanne}, {Switzerland}, {September} 5–10, 2021, {Proceedings}, {Part} {I}}}. \bibinfo{publisher}{Springer-Verlag}, \bibinfo{address}{Berlin, Heidelberg}, \bibinfo{pages}{131--146}.
\newblock
\showISBNx{978-3-030-86548-1}
\href{https://doi.org/10.1007/978-3-030-86549-8_9}{doi:\nolinkurl{10.1007/978-3-030-86549-8_9}}


\bibitem[Shi et~al\mbox{.}(2025)]%
        {shi_augmenting_2025}
\bibfield{author}{\bibinfo{person}{Yao Shi}, \bibinfo{person}{Boyan Li}, \bibinfo{person}{Yuyu Luo}, \bibinfo{person}{Lei Chen}, {and} \bibinfo{person}{Nan Tang}.} \bibinfo{year}{2025}\natexlab{}.
\newblock \showarticletitle{Augmenting {Realistic} {Charts} with {Virtual} {Overlays}}. In \bibinfo{booktitle}{\emph{Proceedings of the 2025 {CHI} {Conference} on {Human} {Factors} in {Computing} {Systems}}} \emph{(\bibinfo{series}{{CHI} '25})}. \bibinfo{publisher}{Association for Computing Machinery}, \bibinfo{address}{New York, NY, USA}, \bibinfo{pages}{1--23}.
\newblock
\showISBNx{979-8-4007-1394-1}
\href{https://doi.org/10.1145/3706598.3714320}{doi:\nolinkurl{10.1145/3706598.3714320}}


\bibitem[Stasko(1997)]%
        {stasko_using_1997}
\bibfield{author}{\bibinfo{person}{John~T. Stasko}.} \bibinfo{year}{1997}\natexlab{}.
\newblock \showarticletitle{Using student-built algorithm animations as learning aids}. In \bibinfo{booktitle}{\emph{Proceedings of the twenty-eighth {SIGCSE} technical symposium on {Computer} science education}} \emph{(\bibinfo{series}{{SIGCSE} '97})}. \bibinfo{publisher}{Association for Computing Machinery}, \bibinfo{address}{New York, NY, USA}, \bibinfo{pages}{25--29}.
\newblock
\showISBNx{978-0-89791-889-3}
\href{https://doi.org/10.1145/268084.268091}{doi:\nolinkurl{10.1145/268084.268091}}


\bibitem[Subramonyam et~al\mbox{.}(2020)]%
        {subramonyam_texsketch_2020}
\bibfield{author}{\bibinfo{person}{Hariharan Subramonyam}, \bibinfo{person}{Colleen Seifert}, \bibinfo{person}{Priti Shah}, {and} \bibinfo{person}{Eytan Adar}.} \bibinfo{year}{2020}\natexlab{}.
\newblock \showarticletitle{{texSketch}: {Active} {Diagramming} through {Pen}-and-{Ink} {Annotations}}. In \bibinfo{booktitle}{\emph{Proceedings of the 2020 {CHI} {Conference} on {Human} {Factors} in {Computing} {Systems}}} \emph{(\bibinfo{series}{{CHI} '20})}. \bibinfo{publisher}{Association for Computing Machinery}, \bibinfo{address}{New York, NY, USA}, \bibinfo{pages}{1--13}.
\newblock
\showISBNx{978-1-4503-6708-0}
\href{https://doi.org/10.1145/3313831.3376155}{doi:\nolinkurl{10.1145/3313831.3376155}}


\bibitem[Tian et~al\mbox{.}(2025)]%
        {tian_chartgpt_2025}
\bibfield{author}{\bibinfo{person}{Yuan Tian}, \bibinfo{person}{Weiwei Cui}, \bibinfo{person}{Dazhen Deng}, \bibinfo{person}{Xinjing Yi}, \bibinfo{person}{Yurun Yang}, \bibinfo{person}{Haidong Zhang}, {and} \bibinfo{person}{Yingcai Wu}.} \bibinfo{year}{2025}\natexlab{}.
\newblock \showarticletitle{{ChartGPT}: {Leveraging} {LLMs} to {Generate} {Charts} {From} {Abstract} {Natural} {Language}}.
\newblock \bibinfo{journal}{\emph{IEEE Transactions on Visualization and Computer Graphics}} \bibinfo{volume}{31}, \bibinfo{number}{3} (\bibinfo{date}{March} \bibinfo{year}{2025}), \bibinfo{pages}{1731--1745}.
\newblock
\showISSN{1941-0506}
\href{https://doi.org/10.1109/TVCG.2024.3368621}{doi:\nolinkurl{10.1109/TVCG.2024.3368621}}


\bibitem[{Tobias Nöthlich}(2020)]%
        {tobias_nothlich_algovisio_2020}
\bibfield{author}{\bibinfo{person}{{Tobias Nöthlich}}.} \bibinfo{year}{2020}\natexlab{}.
\newblock \bibinfo{title}{{AlgoVis}.io - {Online} {Algorithm} {Visualization}}.
\newblock
\urldef\tempurl%
\url{https://tobinatore.github.io/algovis/}
\showURL{%
\tempurl}


\bibitem[Urquiza-Fuentes and Velázquez-Iturbide(2009)]%
        {urquiza-fuentes_survey_2009}
\bibfield{author}{\bibinfo{person}{Jaime Urquiza-Fuentes} {and} \bibinfo{person}{J.~Ángel Velázquez-Iturbide}.} \bibinfo{year}{2009}\natexlab{}.
\newblock \showarticletitle{A {Survey} of {Successful} {Evaluations} of {Program} {Visualization} and {Algorithm} {Animation} {Systems}}.
\newblock \bibinfo{journal}{\emph{ACM Trans. Comput. Educ.}} \bibinfo{volume}{9}, \bibinfo{number}{2} (\bibinfo{date}{June} \bibinfo{year}{2009}), \bibinfo{pages}{9:1--9:21}.
\newblock
\href{https://doi.org/10.1145/1538234.1538236}{doi:\nolinkurl{10.1145/1538234.1538236}}


\bibitem[Vaithilingam et~al\mbox{.}(2024)]%
        {vaithilingam_dynavis_2024}
\bibfield{author}{\bibinfo{person}{Priyan Vaithilingam}, \bibinfo{person}{{https://orcid.org/0000-0001-6730-5683}}, \bibinfo{person}{{View Profile}}, \bibinfo{person}{Elena~L. Glassman}, \bibinfo{person}{{https://orcid.org/0000-0001-5178-3496}}, \bibinfo{person}{{View Profile}}, \bibinfo{person}{Jeevana~Priya Inala}, \bibinfo{person}{{https://orcid.org/0000-0003-1843-589X}}, \bibinfo{person}{{View Profile}}, \bibinfo{person}{Chenglong Wang}, \bibinfo{person}{{https://orcid.org/0000-0002-5933-6620}}, {and} \bibinfo{person}{{View Profile}}.} \bibinfo{year}{2024}\natexlab{}.
\newblock \showarticletitle{{DynaVis}: {Dynamically} {Synthesized} {UI} {Widgets} for {Visualization} {Editing}}.
\newblock \bibinfo{journal}{\emph{Proceedings of the 2024 CHI Conference on Human Factors in Computing Systems}} (\bibinfo{date}{May} \bibinfo{year}{2024}), \bibinfo{pages}{1--17}.
\newblock
\showISSN{9798400703300}
\href{https://doi.org/10.1145/3613904.3642639}{doi:\nolinkurl{10.1145/3613904.3642639}}


\bibitem[Van~Wyk(2024)]%
        {van2024chatgpt}
\bibfield{author}{\bibinfo{person}{Micheal Van~Wyk}.} \bibinfo{year}{2024}\natexlab{}.
\newblock \showarticletitle{Is ChatGPT an opportunity or a threat?: Preventive strategies employed by academics related to a GenAI-based LLM at a faculty of education}.
\newblock \bibinfo{journal}{\emph{Journal of Applied Learning \& Teaching}} \bibinfo{volume}{7}, \bibinfo{number}{1} (\bibinfo{year}{2024}), \bibinfo{pages}{35--45}.
\newblock


\bibitem[Xiao et~al\mbox{.}(2024)]%
        {xiao_let_2024}
\bibfield{author}{\bibinfo{person}{Shishi Xiao}, \bibinfo{person}{Suizi Huang}, \bibinfo{person}{Yue Lin}, \bibinfo{person}{Yilin Ye}, {and} \bibinfo{person}{Wei Zeng}.} \bibinfo{year}{2024}\natexlab{}.
\newblock \showarticletitle{Let the {Chart} {Spark}: {Embedding} {Semantic} {Context} into {Chart} with {Text}-to-{Image} {Generative} {Model}}.
\newblock \bibinfo{journal}{\emph{IEEE Transactions on Visualization and Computer Graphics}} \bibinfo{volume}{30}, \bibinfo{number}{1} (\bibinfo{date}{Jan.} \bibinfo{year}{2024}), \bibinfo{pages}{284--294}.
\newblock
\showISSN{1941-0506}
\href{https://doi.org/10.1109/TVCG.2023.3326913}{doi:\nolinkurl{10.1109/TVCG.2023.3326913}}


\bibitem[Zhou et~al\mbox{.}(2024)]%
        {zhou_epigraphics_2024}
\bibfield{author}{\bibinfo{person}{Tongyu Zhou}, \bibinfo{person}{Jeff Huang}, {and} \bibinfo{person}{Gromit Yeuk-Yin Chan}.} \bibinfo{year}{2024}\natexlab{}.
\newblock \showarticletitle{Epigraphics: {Message}-{Driven} {Infographics} {Authoring}}. In \bibinfo{booktitle}{\emph{Proceedings of the 2024 {CHI} {Conference} on {Human} {Factors} in {Computing} {Systems}}} \emph{(\bibinfo{series}{{CHI} '24})}. \bibinfo{publisher}{Association for Computing Machinery}, \bibinfo{address}{New York, NY, USA}, \bibinfo{pages}{1--18}.
\newblock
\showISBNx{979-8-4007-0330-0}
\href{https://doi.org/10.1145/3613904.3642172}{doi:\nolinkurl{10.1145/3613904.3642172}}


\bibitem[Šimoňák(2020)]%
        {simonak_increasing_2020}
\bibfield{author}{\bibinfo{person}{Slavomír Šimoňák}.} \bibinfo{year}{2020}\natexlab{}.
\newblock \showarticletitle{Increasing the {Engagement} {Level} in {Algorithms} and {Data} {Structures} {Course} by {Driving} {Algorithm} {Visualizations}}.
\newblock \bibinfo{journal}{\emph{Informatica}} \bibinfo{volume}{44}, \bibinfo{number}{3} (\bibinfo{date}{Sept.} \bibinfo{year}{2020}).
\newblock
\showISSN{1854-3871}
\href{https://doi.org/10.31449/inf.v44i3.2864}{doi:\nolinkurl{10.31449/inf.v44i3.2864}}


\end{thebibliography}

\crefalias{section}{appendix}
\crefalias{subsection}{appendix}
\clearpage
\appendix
\section{Prompts for AV Code Generator}
In each prompt, certain variables are enclosed within curly brackets and colored blue for easy recognition.

\subsection{Prompt for Pseudocode Analysis}
The following variables are used in this prompt:
\begin{itemize}
    \item \texttt{TIME\_SPACE\_COMPLEXITY\_OPTIONS}: Available options for time and space complexity analysis.
    \item \texttt{state["pseudocode"]}: Text extracted from the pseudocode image using OCR.
\end{itemize}

\begin{lstlisting}[style=codebox]
You are an expert in data structures and algorithms.

# Task
Analyze the algorithm name, input arguments, worst-case time complexity, and space complexity from the pseudocode.

# Input
OCR result of pseudocode: {state["pseudocode"]}

# Output
Concise and structured output containing:
- Algorithm name
- Input arguments
- Worst-case Time complexity: Choose from {TIME_SPACE_COMPLEXITY_OPTIONS}
- Space complexity: Choose from {TIME_SPACE_COMPLEXITY_OPTIONS}
\end{lstlisting}

\subsection{Prompts for Algorithm Code Generation}
The following variables are used in this prompt:
\begin{itemize}
    \item \texttt{state["programming\_language"]}: Programming language for implementation (Python or JavaScript).
    \item \texttt{state["pseudocode"]}: Text extracted from the pseudocode image using OCR.
    \item \texttt{state["executable\_algo\_code"]}: Generated executable algorithm code.
    \item \texttt{state["algo\_error\_message"]}: Execution or interpreter error message.
\end{itemize}

\subsubsection{Algorithm Code Generation}
\leavevmode
\begin{lstlisting}[style=codebox]
You are an expert in {state["programming_language"]} code generation.

# Task
Generate clean, idiomatic, and **fully executable** {state["programming_language"]} code from the pseudocode.

# Input
Pseudocode: {state["pseudocode"]}

# Output
Raw text of the translated code.

## Example:
---
# @line(0) BUBBLESORT(A)
def bubble_sort(A):
    # @line(1) for i = 0 to A.length - 1
    for i in range(len(A)):
        ...

if __name__ == "__main__":
    bubble_sort([5, 3, 8, 4, 2])
---

## Strict Formatting Rules:
- Output ONLY the raw code text; do NOT include markdown blocks or explanations.
- Each pseudocode line that is translated must have an `@line(x)` comment immediately before the corresponding code, where `x` is the zero-based index of the line in the original pseudocode (including comments and non-code lines).
- Include all function definitions and ensure the program is complete (imports + main/test code).
- Maintain the logic of the original pseudocode exactly.
- Use clear and idiomatic {state["programming_language"]} conventions.
- Include example usage such as `if __name__ == "__main__"` or sample input/output.
\end{lstlisting}

\subsubsection{Reflection}
\leavevmode
\begin{lstlisting}[style=codebox]
You are an expert in {state["programming_language"]} code generation.

# Task
Fix the executable algorithm code to make it run successfully.

# Input
- Executable algorithm code: {state["executable_algo_code"]}
- Error message: {state["algo_error_message"]}

# Output
Raw text of the fixed algorithm executable code

## Strict Formatting Rules:
- Output ONLY the raw code text; do NOT include markdown blocks or explanations.
\end{lstlisting}

\subsection{Prompt for Visualization Type Selection}
The following variables are used in this prompt:
\begin{itemize}
    \item \texttt{VISUALIZATION\_METHODS}: Available visualization methods (1D Array, Bar Chart, Linked List, and Graph).
    \item \texttt{state["pseudocode"]}: Text extracted from the pseudocode image using OCR.
    \item \texttt{state["executable\_algo\_code"]}: Generated executable algorithm code.
\end{itemize}
\begin{lstlisting}[style=codebox]
You are an expert in algorithm visualization.

# Task
Analyze the pseudocode and executable algorithm code, and select the most appropriate visualization method(s) from the list provided. You must consider the characteristics of the algorithm (e.g., data structure used, traversal style, state changes) and the logic shown in the pseudocode.

# Input
- Pseudocode: {state["pseudocode"]}
- Executable algorithm code: {state["executable_algo_code"]}

# Output
Concise and structured output containing:
- Selected Visualization Method(s): List the most suitable visualization methods from {VISUALIZATION_METHODS} (you may choose multiple).
\end{lstlisting}

\subsection{Prompts for Visualization Command Injection}
The following variables are used in this prompt:
\begin{itemize}
    \item \texttt{TEMPLATE\_CODE["programming\_language"]}: Template for algorithm visualization in each programming language.
    \item \texttt{visualization\_mapping\_string}: Mapping of visualization methods to class names.
    \item \texttt{vis\_lib\_cheat\_sheet}: Concise documentation for selected visualization methods.
    \item \texttt{state["programming\_language"]}: Programming language for implementation (Python or JavaScript).
    \item \texttt{state["executable\_algo\_code"]}: Generated executable algorithm code.
    \item \texttt{state["executable\_vis\_code"]}: Generated algorithm visualization code.
    \item \texttt{state["vis\_error\_message"]}: Execution or interpreter error messages.
    
\end{itemize}

\subsubsection{Visualization Command Injection}
\leavevmode
\begin{lstlisting}[style=codebox]
You are an expert in {state["programming_language"]} code generation.

# Task
Fill in the `<imports>` and `<stub>` sections of {TEMPLATE_CODE[state["programming_language"]]} and generate clean, idiomatic, and **fully executable** {state["programming_language"]} code to visualize interesting events of a given algorithm.

# Input
- Executable algorithm code: {state["executable_algo_code"]}
- Available visualization classes and methods: {visualization_mapping_string}
- API cheat sheet: {vis_lib_cheat_sheet}

# Output
Raw text of the generated algorithm visualization code.

## Strict Formatting Rules:
- Output ONLY the raw code text; do NOT include markdown blocks or explanations.
- EVERY `@line` comment from the original algorithm code MUST be immediately followed by a structural `logger.println()` call, as detailed in Section 3. Traceability Logging.
- The main algorithm function must be called directly at the end of the script. DO NOT use an `if __name__ == '__main__':` block.
- Your final output must exactly match this structure, 
- You MUST NOT alter the logic of the provided algorithm code in any way.
- Enclose all added code (visualization, highlighting, traceability logging, delay) within \textcolor{black}{`# {{` and `# }}` comments. This is mandatory for parsing.

## Implementation Guide:

### 1. Input Data Generation
- Input data MUST be dynamically created using the `Randomize` class. Do not use static examples.
- Using the generated input data, provide one most important scenario for the algorithm. This should be a single, meaningful input that showcases the algorithm's behavior effectively.

### 2. Visualization & Highlighting
- Drawing: Use the specified visualization classes and methods to render the data structures.
- Chart Sync: If using both `ChartTracer` and `Array1DTracer`, you MUST call `Array1DTracer.chart()` to sync them. Do not use `ChartTracer.set()` in this scenario.
- Highlighting States (Mutually Exclusive):
    - `select`: Use when **comparing** elements.
    - `patch`: Use **just before** an element's value is changed or swapped.
    - `lock`: Use when an element becomes part of the **final, sorted partition** of the array.
- Logic for Applying `lock`:
    - Identify the Sorted Partition: Most sorting algorithms work by progressively building a "sorted partition" (e.g., at the beginning or end of the array) and shrinking the "unsorted partition."
    - Find the Newly Finalized Element: After each full pass of the algorithm's main outer loop (e.g., the `i` loop), your task is to identify which **one** new element has just been placed into its final, permanent position, thus expanding the sorted partition.
    - Apply the Lock: Call the `lock()` method on the index of that single, newly finalized element. For example, after the first pass of the outer loop, one element will be sorted. After the second pass, a second element will be sorted, and so on.
- State Management: Before applying a new state (`select`, `patch`, `lock`), you MUST clear any pre-existing state from the element (e.g., call `deselect()` before `patch()`).
- Pausing: Call `Tracer.delay()` after each logical step to create a visual pause.
- Example (Swapping two elements):
     When swapping `A[j]` and `A[j-1]`, you must call `patch` for **both** elements with their **new, future values** before the actual swap.
    ---
    # @line(4) exchange A[j] with A[j - 1]
    # {{
    logger.println(f"@line(4) exchange A[{{j}}] @pos(from=A[j], to={{A[j]}}) with A[{{j - 1}}] @pos(from=A[j-1], to={{A[j - 1]}})")
    # }}
    # 1. Announce the upcoming changes with patch()
    array_tracer.patch(j, A[j - 1]); # Announce that A[j] will become the value of A[j-1]
    array_tracer.patch(j - 1, A[j]); # Announce that A[j-1] will become the value of A[j]
    Tracer.delay();

    # 2. Perform the actual swap
    const temp = A[j];
    A[j] = A[j - 1];
    A[j - 1] = temp;
    
    # 3. Optional: Clear the patch highlight after the swap is done
    array_tracer.depatch(j);
    array_tracer.depatch(j-1);
    ---

### 3. Traceability Logging
- Core Principle: The goal is a **one-to-one mapping**. Each pseudocode function arguments, variable, value, or expression MUST have its own, separate `@pos` token.
- Crucial Rule: **NEVER group multiple mappings into a single `@pos` token**, especially for ranges (like `start..end`).
- Format: Use `@pos(from=<pseudo_value>, to=<impl_value>)` for **each individual mapping**.
- Example (Handling Ranges):
    - Pseudocode Line: `# @line(2) for j = A.length downto i + 1`
    - INCORRECT (Grouping values):
    logger.println(f"... @pos(from=A.length..i+1, to={{len(A)-1}}..{{i+1}})")
    - CORRECT (Separate tokens for each value):
    logger.println(f"... @pos(from=A.length, to={{len(A)-1}}) downto ... @pos(from=i+1, to={{i+1}})")
    - Explanation: Notice how `A.length` and `i + 1` each have their own independent `@pos` token in the correct version. This is mandatory.
\end{lstlisting}

\subsubsection{Reflection}
\leavevmode
\begin{lstlisting}[style=codebox]
You are an expert in {state["programming_language"]} code generation.

# Task
Fix the algorithm visualization code to make it run successfully.

# Input
- Algorithm visualization code: {state["executable_vis_code"]}
- Error message: {state["vis_error_message"]}

# Output
Raw text of the fixed algorithm visualization code

## Strict Formatting Rules:
- Output ONLY the raw code text; do NOT include markdown blocks or explanations.
\end{lstlisting}

\section{Usability Study Participant Demographics}
\label{appdix:demographics}
\begin{table}[H]
\centering
\small
\begin{tabular}{lllll}
\toprule
ID & Age & Gender & Education & DSA Familiarity$^1$ \\
\midrule
P1  & 23 & Female & Postgraduate & 6 \\
P2  & 23 & Female & Postgraduate & 6 \\
P3  & 20 & Female & Bachelor     & 5 \\
P4  & 22 & Female & Postgraduate & 6 \\
P5  & 22 & Male   & Bachelor     & 5 \\
P6  & 21 & Male   & Bachelor     & 5 \\
P7  & 23 & Female & Postgraduate & 5 \\
P8  & 25 & Female & Postgraduate & 7 \\
P9  & 25 & Male   & Postgraduate & 6 \\
P10 & 21 & Female & Bachelor     & 5 \\
P11 & 21 & Female & Bachelor     & 5 \\
P12 & 23 & Female & Postgraduate & 4 \\
P13 & 21 & Male   & Bachelor     & 6 \\
P14 & 21 & Female & Bachelor     & 5 \\
P15 & 22 & Male   & Bachelor     & 6 \\
\bottomrule
\multicolumn{5}{l}{$^1$ 1=Low, 7=High.}
\end{tabular}
\caption{Participant Demographics}
\label{tab:participants}
\end{table}

\section{Usability Study System Usability Score}
\label{appdix:sus}
\begin{figure*}
\centerline{\includegraphics[width=.9\linewidth]{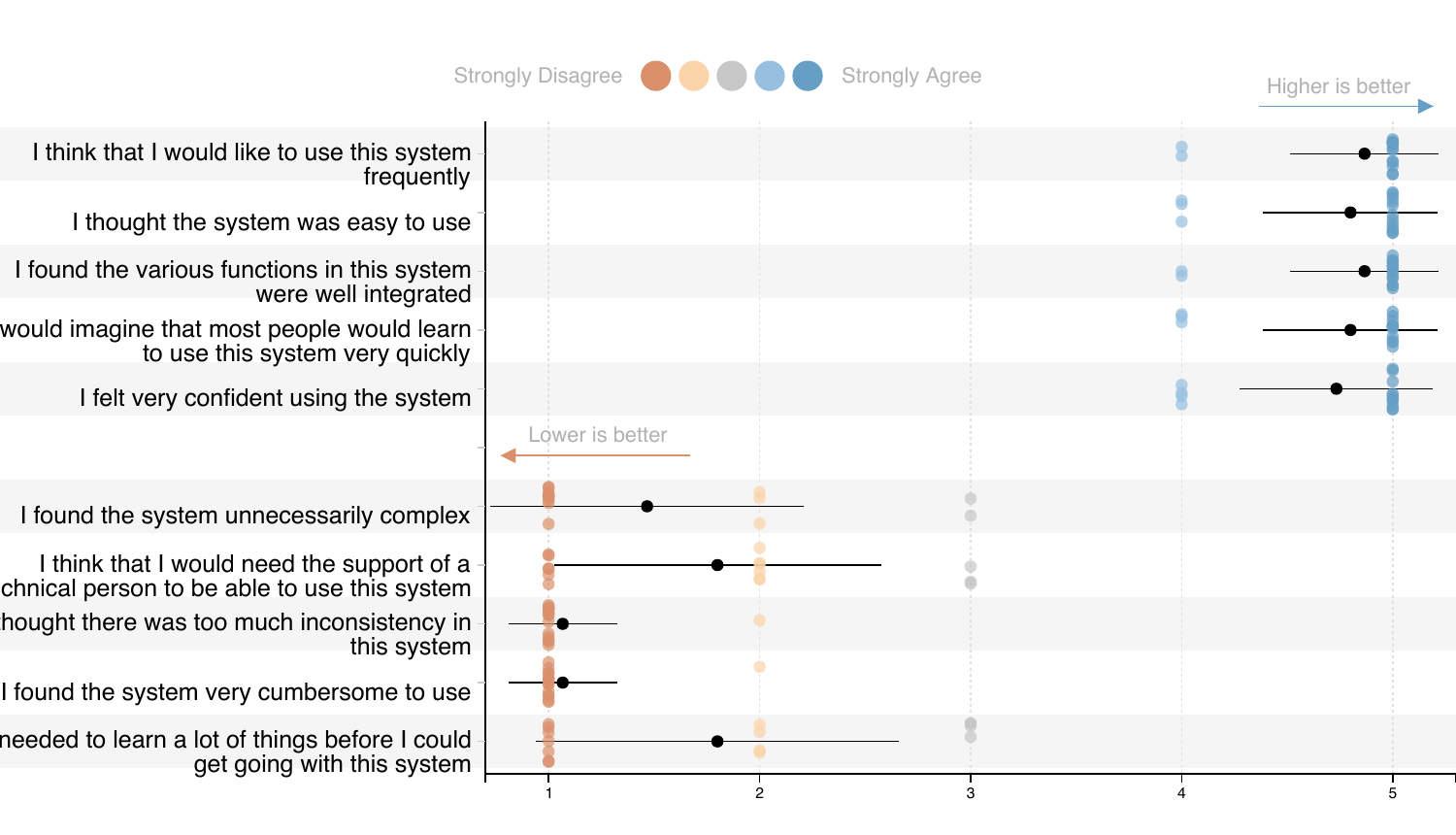}}
\caption{Mean and standard deviation of participants’ answers to the SUS. Questions are divided into positive items (top half, higher is better) and negative items (bottom half, lower is better).
}
\label{fig:sus-result}
\end{figure*}

\end{document}